\documentclass[11pt,a4paper]{article}
\usepackage{jheppub}
\usepackage{amsmath}
\usepackage{amssymb}
\usepackage{stmaryrd}
\DeclareMathOperator{\PP}{\mathbb{P}}
\DeclareMathOperator{\BPP}{\bar{\mathbb{P}}}

\newtheorem{remark}{Remark}

\title{Investigations into Light-front Interactions for Massless Fields (I): Non-constructibility of Higher Spin Quartic Amplitudes}
\author[a,1]{Anders K. H. Bengtsson\note{Work supported by the Research and Education Board at the University of Bor{\aa}s.}}

\affiliation[a]{Academy of Textiles, Engineering and Economics, University of Bor{\aa}s, All\' egatan 1, SE-50190 Bor\aa s, Sweden.}

\emailAdd{anders.bengtsson@hb.se} 

\abstract{The dynamical commutators of the light-front Poincar{\'e} algebra yield first order differential equations in the $p^+$ momenta for the interaction vertex operators. The homogeneous solution to the equation for the quartic vertex is studied. Consequences as regards the constructibility assumption of quartic higher spin amplitudes from cubic amplitudes are discussed. The existence of quartic contact interactions unrelated to cubic interactions by Poincar{\'e} symmetry indicates that the higher spin S-matrix is not constructible. Thus quartic amplitude based no-go results derived by BCFW recursion for Minkowski higher spin massless fields may be circumvented.}


\begin{document}

\maketitle
\pagebreak

\section{Introduction}\label{sec:Introduction}
Reproducibility is a basic tenet of science -- in particular as regards experimental results -- but it also applies to theoretical subjects. In fields were many researchers work, this is no great problem. However, there are areas of theoretical physics which are not that populated. The theory of massless higher helicity fields in Minkowski light-front space-time is one such area. This is so even though higher spin gauge field theory itself has become a very active subject.\footnote{There indeed seems to be a renewed interest in Minkowski space higher spin theory as evidenced for instance by recent work \cite{PonomarevTseytlin2016a,CondeJoungMkrtchyan2016,Ponomarev2016a}.}

In the early 1990's, R. Metsaev studied quartic interactions for higher spin fields \cite{MetsaevQuartic1,MetsaevQuartic2} in the light-front gauge. As I have been myself interested in that problem for a long time I have decided to see if I can reproduce Metsaev's results and perhaps extend on them, using the momentum space vertex operator approach of our 1987 paper \cite{BBL1987}. Based on systematisation and extension of the results of that paper -- done in \cite{AKHB2012a} and \cite{AKHB2016a} -- the present paper offers a first step in the study of the general quartic vertex operator.\footnote{As far as I'm aware of, there are other groups  working on this problem at the present time. This is very good and hopefully we will be able to settle the question (through independent but related efforts) about the existence of consistent quartic interactions in the near future.}

As regards the contents of the Metsaev papers, there are two main results (apart from a formal solution for the quartic interaction itself): First, that there are quartic vertices that are Poincar{\'e} invariant by themselves independently of the cubic vertices. Second, that the coefficients for the cubic vertices are determined to have a certain form by Poincar{\'e} invariance at the quartic level. The first result we will be able to verify in the present paper. The second result will be investigated in a subsequent paper. A detailed term-by-term comparison is planned for a separate work. 

One further object of the present paper is to \emph{understand} the very structure of the light-front vertex computations. Due to the complexity of the problem, such understanding will presumably be critical to an attempt at the quintic order. Of course, after that, proceeding order by order is likely to be too hard and not very illuminating. The more interesting goal is an all orders existence proof of Minkowski light-front higher spin interactions. \emph{Such a proof -- if it indeed can be constructed -- must rely on generic properties of the deformation equations for the vertices and deep understanding of what the equations mean.}

This is the plan: Section \ref{sec:FreeFields} sets up the free field theory (for more details, see \cite{AKHB2012a}). In section \ref{sec:Kinematics} some very useful results on cubic and quartic momentum kinematics is reviewed (for more details, see \cite{AKHB2016a}). The quartic kinematics is crucial for the efficient study of the interactions and seems not to have been explicitly stated before. Section \ref{sec:Interactions} sets up the general scheme for light-front interactions (for more details, see \cite{AKHB2012a}). Section \ref{sec:ReviewCubic} derives the well known cubic vertex operator, but in a streamlined form amenable to generalisation to higher orders. In section \ref{sec:HomogenousQuartic} then, the homogeneous part of the quartic vertex operator is studied. 

The terminology \emph{homogeneous} stems from the fact that the equations determining the vertex operator of order $\nu$ are first order differential equations in the light-front $p^+$ momentum with right hand sides given by lower order vertices. In this paper we only study the solution to the corresponding homogeneous equation. This means that we are studying quartic interactions that are independent of the cubic interactions.\footnote{Such interactions have been previously studied by M. Taronna in a covariant formalism in \cite{Taronna2011a}.} The calculations needed are spelled out in detail (which has not been done before) in order to better understand them. In section \ref{sec:QuarticVertices}, the generic structure of quartic vertices will be discussed and examples of homogeneous contact interactions will be given.

Consequences of the existence of solutions to the homogeneous equation are discussed in section \ref{sec:Consequences}. The constructibility assumption -- in the sense that quartic amplitudes can be reconstructed from the cubic interaction vertices only -- made in BCFW derivations of the quartic higher spin amplitude \cite{BenincasaCachazo2008a,BenincasaConde2011a} (for a review from a light-front perspective, see also \cite{AKHB2016b}) may not apply.\footnote{BCFW constructibility of flat space higher spin theory in relation to quartic interactions has also been discussed in \cite{FotopoulosTsulaia2010a,DempsterTsulaia2012a} based on  \cite{BenincasaCachazo2008a,BenincasaConde2011a}.} Provided higher order consistency does not rule out the new quartic vertices, their presence may circumvent BCFW based no-go arguments as to the possibility of a consistent Minkowski space-time higher spin theory. The question of constructibility has also been recently discussed within a concrete tree level calculation of a scalar four-point amplitude \cite{PonomarevTseytlin2016a} with a tower of higher spin particles in the channel. For further comments on this, see section \ref{sec:Consequences}.\footnote{It should perhaps be pointed out that the considerations in the present paper are entirely within classical field theory. We are searching for classical interaction terms contributing to a Poincar{\'e} invariant action for a higher spin theory formulated on the Minkowski light-front. Inconsistencies may well turn up in the corresponding quantum theory, even though the vertices found here may evade one particular kind of inconsistency.}

Let me end the introduction with a comment about the philosophy guiding this work. Throughout the text, I have inserted some \emph{understanding remarks}. It has for a long time been my opinion that the problem of constructing free-standing higher spin theories (as opposed to when higher spins occur as excitations in other models) is to a large extent a problem of controlling the inherent complexity of the problem (see my paper \cite{AKHB2005a}). In AdS spaces this is achieved by the Vasiliev equations \cite{Vasiliev1999Review,Vasiliev2004Reviewa,Vasiliev2004Reviewb,Vasiliev2005a,BekaertCnockaertIazeollaVasiliev2005}, but in Minkowski space-time we have nothing of that strength. It could perhaps be speculated that in order to solve the problem one would need to go outside the context of set-based mathematics and work in some more general category \cite{NewStructuresPhysics} in order to dissolve the problem by raising the level of abstraction. But as physicists, we still want to make computations and get real and complex functions out. 

On the other hand, the $D=4$ light-cone is as concrete as it gets. Working with only physical fields is a great simplification, especially in four dimensions where there are just the two helicities for each and every spin. The backside is the non-covariance of the formalism that easily produces formulas that are hard to handle. The \emph{understanding remarks} are meant to highlight what is in my opinion important (although often simple) insights into to workings of the light-front formalism. Looked upon in the right way, the light-front may be more transparent than it seems at first. And as some workers has noted (in particular S. Ananth \cite{Ananth2012un}), the light-front formalism is very close to the spinor helicity formalism (further explored in a recent paper \cite{Ponomarev2016a}) and to twistor theory \cite{HaehnelMcLoughlin2016a}, furthermore hinting at an underlying simplicity that we just haven't uncovered yet.

\section{Free light-front higher helicity fields}\label{sec:FreeFields}

Fields of all integer helicities can be collected in a Fock space field
\begin{equation}\label{eq:FockSpaceField}
\vert\Phi(p)\rangle=\sum_{\lambda=0}^\infty\frac{1}{\sqrt{\lambda!}}\left(\phi_\lambda(p)(\bar\alpha^\dagger)^\lambda+\bar{\phi}_\lambda(p)(\alpha^\dagger)^\lambda\right)\vert 0\rangle
\end{equation}
where $p$ is short for $p,\bar{p}$ and $\gamma=p^+$. This Fock space field is real in the sense that
\begin{equation}\label{eq:RealityFockField}
\begin{split}
\vert\Phi(p)\rangle^\dagger&=\langle\Phi(p)\vert\\
&=\sum_{\lambda=0}^\infty\frac{1}{\sqrt{\lambda!}}\langle 0\vert\left(\bar{\phi}_\lambda(p)(\alpha)^\lambda+\phi_\lambda(p)(\bar{\alpha})^\lambda\right)
\end{split}
\end{equation}
The two-dimensional complex internal Fock space is spanned by oscillators $\alpha^\dagger$ and $\bar\alpha^\dagger$ where
\begin{equation}\label{eq:OscillatorCommutators}
[\alpha,\bar\alpha^\dagger]=[\bar\alpha,\alpha^\dagger]=1
\end{equation}
In the interacting theory, a shorthand notation is used: $\vert\Phi_r\rangle$ stands for $\vert\Phi(\gamma_r,p_r,\bar{p}_r)\rangle$ expanded over $\alpha_r^\dagger$ and $\bar{\alpha}_r^\dagger$ and correspondingly for $\langle\Phi_r\vert$. In an $\nu$-order interaction term, $r$ will run from $1$ to $\nu$ and serve as a label on the fields. The field \eqref{eq:FockSpaceField} is subject to the tracelessness constraint $T|\Phi\rangle=0$ with $T=\bar\alpha\alpha$ that prevents mixed excitations such as $\alpha^\dagger\bar{\alpha}^\dagger$. However, the theory works just as well if such excitations are allowed (for details, consult \cite{AKHB2012a}). The free theory Hamiltonian is
\begin{equation}\label{eq:FreeHamiltonian}
H_{(0)}=\frac{1}{2}\int \gamma d\gamma dpd\bar{p}\langle\Phi(p)|h|\Phi(p)\rangle\quad\text{with}\quad h=\frac{p\bar{p}}{\gamma}\quad\text{and}\quad \gamma=p^+
\end{equation}
%

\section{Kinematics}\label{sec:Kinematics}
The kinematics of the interactions are such that for an order $\nu$ interaction we have momentum conservation
\begin{equation}\label{eq:Momentumonservation}
\sum_{r=1}^\nu p_r=\sum_{r=1}^\nu \bar{p}_r=\sum_{r=1}^\nu \gamma_r=0
\end{equation}
It is convenient to write the transverse momentum dependence in terms of the combinations
\begin{equation}\label{eq:PPandBPPtransversemomenta}
\PP_{ij}=\gamma_ip_j-\gamma_jp_i\quad\text{and}\quad \BPP_{ij}=\gamma_i\bar{p}_j-\gamma_j\bar{p}_i
\end{equation}
The number of $\PP_{ij}$ for an order $\nu$ vertex is $n=\nu(\nu-1)/2$. Due to momentum conservation, only $n-2$ of those are linearly independent. For the cubic, $\nu=3$, this means that there is only one $\PP=\PP_{12}=\PP_{23}=\PP_{31}$ and similarly for $\BPP$.

Based on this it is possible to derive linear recombination formulas. Let $c_r$ be arbitrary variables, then we have for the cubic
\begin{equation}\label{eq:recombinationFormulaCubic}
\sum_{r=1}^3 c_rp_r=\frac{1}{3}\Big(\sum_{r=1}^3 c_r\gamma_r\Big)\Big(\sum_{s=1}^3\frac{p_s}{\gamma_s}\Big)-\frac{1}{3}\Big(\sum_{r=1}^3 S_rc_r\Big)\PP
\end{equation}
where $S_r=1/\gamma_{r+1}-1/\gamma_{r+2}$. In this formula, the objects
\begin{equation}\label{eq:NewCubicBasis}
\sum_{s=1}^3\frac{p_s}{\gamma_s}\quad\text{and}\quad \PP
\end{equation}
are independent basis vectors in the two-dimensional transverse momentum space $(p_1,p_2,p_3)$.

For the quartic \cite{AKHB2016a} we have in the $\{\PP_{12},\PP_{34}\}$ basis (corresponding to the $\mathbf{s}$-channel) 
\begin{equation}\label{eq:recombinationFormulaQuartic3Schannel}
\sum_{r=1}^4 c_rp_r=\frac{1}{4}\Big(\sum_{r=1}^4 c_r\gamma_r\Big)\Big(\sum_{s=1}^4\frac{p_s}{\gamma_s}\Big)-\frac{1}{4}\left(\sum_{r=1}^4S_{12,r}c_r\right)\PP_{12}-\frac{1}{4}\left(\sum_{r=1}^4S_{34,r}c_r\right)\PP_{34}
\end{equation}
In this formula, the objects
\begin{equation}\label{eq:NewQuarticBasis}
\sum_{s=1}^4\frac{p_s}{\gamma_s},\; \PP_{12}\;\text{and}\; \PP_{34}
\end{equation}
are independent basis vectors in the three-dimensional transverse momentum space $(p_1,p_2,p_3,p_4)$. The coefficients $S_{12,r}$ and $S_{34,r}$ are rational functions of the $\gamma$'s given by
\begin{align}\label{eq:CoefficientsFormulaQuartic3Schannel}
S_{12,1}&=\frac{3\gamma_2+\gamma_1}{\gamma_2(\gamma_1+\gamma_2)} & S_{34,3}&=\frac{3\gamma_4+\gamma_3}{\gamma_4(\gamma_3+\gamma_4)}\\
S_{12,2}&=-\frac{3\gamma_1+\gamma_2}{\gamma_1(\gamma_1+\gamma_2)} & S_{34,4}&=-\frac{3\gamma_3+\gamma_4}{\gamma_3(\gamma_3+\gamma_4)}\\
S_{12,3}&=\frac{\gamma_3(\gamma_1-\gamma_2)}{\gamma_1\gamma_2(\gamma_1+\gamma_2)}& S_{34,1}&=\frac{\gamma_1(\gamma_3-\gamma_4)}{\gamma_3\gamma_4(\gamma_3+\gamma_4)}\\
S_{12,4}&=\frac{\gamma_4(\gamma_1-\gamma_2)}{\gamma_1\gamma_2(\gamma_1+\gamma_2)}& S_{34,2}&=\frac{\gamma_2(\gamma_3-\gamma_4)}{\gamma_3\gamma_4(\gamma_3+\gamma_4)}
\end{align}
The coefficients are listed so that it is easy to see that the formula is symmetric under the interchange of labels $1\leftrightarrow 3$ and $2\leftrightarrow 4$. Similar formulas can be written for $\mathbf{t}$-channel and $\mathbf{u}$-channel variables. 

\begin{remark}[Understanding the kinematics]\label{rem:UKinematics}
The importance of the formulas \eqref{eq:recombinationFormulaCubic} and \eqref{eq:recombinationFormulaQuartic3Schannel} cannot be over stressed. The cubic equation was used in light-front string field theory in the 1980's. The quartic is to the best of my knowledge new, and it generalises to any interaction order \cite{AKHB2016a}. One way of viewing these equations is to see that they allow us to separate the kinematical from the truly dynamical, the dynamics pointing in the $\PP_{ij}$ and $\BPP_{ij}$ directions. The meaning of this comment will be clear as we continue.
\end{remark}

\section{Interactions}\label{sec:Interactions}

The interaction Hamiltonian of order $\nu$ is written as
\begin{equation}\label{eq:OrderNuIntercationFull}
H_{\scriptsize{(\nu-2)}}=\frac{1}{\nu}\int\prod_{r=1}^\nu \gamma_rd\gamma_r dp_rd\bar{p}_r\langle\Phi_r\vert V_{1\ldots\nu}\rangle
\end{equation}
where the $\nu$-th order vertex is
\begin{equation}\label{eq:NuVertexFull}
\vert V_{1\ldots\nu}\rangle=\Big(\frac{g}{\kappa}\Big)^{4-\nu}\exp\Delta_\nu\vert0_{1\ldots\nu}\rangle\Gamma_\nu^{-1}\delta({\textstyle\sum_r}\gamma_r)\delta({\textstyle\sum_r}p_r)\delta({\textstyle\sum_r}\bar{p}_r)
\end{equation}
where $\Gamma_\nu=\gamma_1\gamma_2\cdots\gamma_\nu$. The power of the coupling is determined by dimensional analysis and $g$ has mass dimension $0$ and $\kappa$ mass dimension $-1$. There are further dimensionful factors in the $\Delta_\nu$ operators so that the spin $1$ coupling come out as dimensionless.

In the following, the momentum integrations will be considered to be included as part of the Fock space inner product $\langle\,\vert\,\rangle$. The momentum delta functions and the factor $\Gamma_\nu^{-1}$ will be included in the vacua. This gives the following shorthand expressions, where we also give the dynamical Lorentz generators
\begin{equation}\label{eq:OrderNuIntercationShort}
\begin{split}
H_{\scriptsize{(\nu-2)}}&=\frac{1}{\nu}\prod_{r=1}^\nu\langle\Phi_r\vert V_{1\ldots\nu}\rangle\\
J^-_{\scriptsize{(\nu-2)}}&=\frac{1}{\nu}\prod_{r=1}^\nu\langle\Phi_r\vert\hat{x}_{(\nu)}\vert V_{1\ldots\nu}\rangle\\
\bar{J}^-_{\scriptsize{(\nu-2)}}&=\frac{1}{\nu}\prod_{r=1}^\nu\langle\Phi_r\vert\hat{\bar{x}}_{(\nu)}\vert V_{1\ldots\nu}\rangle\\
\end{split}
\end{equation}
The vertex $\vert V_{1\ldots\nu}\rangle$ and the vacuum $\vert\varnothing_{1\ldots\nu}\rangle$ are given by
\begin{equation}\label{eq:OrderVertexVacuumShort}
\begin{split}
\vert V_{1\ldots\nu}\rangle&=(g\kappa^{-1})^{4-\nu}\exp\Delta_\nu\vert\varnothing_{1\ldots\nu}\rangle\\
\vert\varnothing_{1\ldots\nu}\rangle&=\Gamma_\nu^{-1}\delta({\textstyle\sum_r}\gamma_r)\delta({\textstyle\sum_r}p_r)\delta({\textstyle\sum_r}\bar{p}_r)\vert0_{1\ldots\nu}\rangle
\end{split}
\end{equation}
Interaction data is encoded in the $\Delta_\nu$ and the $\hat{x}_{(\nu)}$ and $\hat{\bar{x}}_{(\nu)}$. The dynamical generators to all orders are now given by
\begin{equation}\label{eq:AllOrdersHamiltonian}
H=H_{(0)}+\sum_{\nu=3}^{\infty} H_{(\nu-2)}
\end{equation}
and similar expressions for the dynamical Lorentz generators $J^-$ and $\bar{J}^-$. The dynamical part of the Poincar{\'e} algebra then yields recursive equations for $\Delta_\nu$ and the $\hat{x}_\nu$ and $\hat{\bar{x}}_\nu$. In trying to solve these equations we need an ansatz. The form of such an ansatz is restricted by the kinematical part of the algebra (see \cite{AKHB2012a} for details).

The operators \eqref{eq:OrderNuIntercationShort} generate transformations according to 
\begin{equation}\label{eq:GeneralTransformation}
\delta_G|\Phi\rangle=[|\Phi\rangle,G]
\end{equation}
The exact workings of the commutator $[\cdot,\cdot]$ can be found in reference \cite{AKHB2012a}. For two generic dynamical generators $A$ and $B$ and a field $\vert\Phi_\chi\rangle$ we have
\begin{equation}\label{eq:AbstractAlgebra}
[\delta_A,\delta_B]\vert\Phi_\chi\rangle=0
\end{equation}
Expanding this equation a few orders in the interaction we get 
\begin{equation}\label{eq:ExpansionOfAlgebraToQuintic}
\begin{split}
&\text{Free:}\hspace{25pt}[\delta_A^{(0)},\delta_B^{(0)}]\vert\Phi_\chi\rangle=0\\
&\text{Cubic:}\hspace{17pt}\big([\delta_A^{(0)},\delta_B^{(1)}]+[\delta_A^{(1)},\delta_B^{(0)}]\big)\vert\Phi_\chi\rangle=0\\
&\text{Quartic:}\hspace{9pt}\big([\delta_A^{(0)},\delta_B^{(2)}]+[\delta_A^{(2)},\delta_B^{(0)}]\big)\vert\Phi_\chi\rangle=-[\delta_A^{(1)},\delta_B^{(1)}]\vert\Phi_\chi\rangle
\end{split}
\end{equation}
The general form of these \emph{deformation equations} can be written as
\begin{equation}\label{eq:ExpansionOfAlgebraGeneral}
\big([\delta_A^{(0)},\delta_B^{(\nu)}]+[\delta_A^{(\nu)},\delta_B^{(0)}]\big)\vert\Phi_\chi\rangle=-\sum_{\mu=1}^{\nu-1}\Big([\delta_A^{(\mu)},\delta_B^{(\nu-\mu)}]+[\delta_A^{(\nu-\mu)},\delta_B^{(\mu)}]\Big)\vert\Phi_\chi\rangle
\end{equation}
The left hand side will be called the {\it differential commutator} and the right hand side the {\it source commutator}. The equations become first order differential equations for the $\gamma_r$ dependence of the $\Delta_\nu$ operators. The differential commutator can be further reduced to a concrete form suitable for calculation. We list the three differential commutators corresponding to the dynamical part of the algebra.
\begin{subequations}\label{eq:DifferentialSide}
\begin{equation}\label{eq:DifferentialSideA}
\sum_{r=1}^\nu j_r^-\vert V_{1\ldots\nu}\rangle-\sum_{r=1}^\nu h_r\hat{x}_{(\nu)}\vert V_{1\ldots\nu}\rangle
\end{equation}
\begin{equation}\label{eq:DifferentialSideB}
\sum_{r=1}^\nu \bar{j}_r^-\vert V_{1\ldots\nu}\rangle-\sum_{r=1}^\nu h_r\hat{\bar{x}}_{(\nu)}\vert V_{1\ldots\nu}\rangle
\end{equation}
\begin{equation}\label{eq:DifferentialSideC}
\sum_{r=1}^\nu \bar{j}_r^-\hat{x}_{(\nu)}\vert V_{1\ldots\nu}\rangle-\sum_{r=1}^\nu j_r^-\hat{\bar{x}}_{(\nu)}\vert V_{1\ldots\nu}\rangle
\end{equation}
\end{subequations}
In these expressions, $j^-$ and $\bar{j}^-$ are the free theory dynamical Lorentz generators
\begin{equation}\label{eq:JminusLorentzGenerators}
\begin{split}
j^-&=xh+ip\frac{\partial}{\partial\gamma}-\frac{i}{\gamma}Mp\\
\bar{j}^-&=\bar{x}h+i\bar{p}\frac{\partial}{\partial\gamma}+\frac{i}{\gamma}M\bar{p}
\end{split}
\end{equation}
where $M=\alpha^\dagger\bar{\alpha}-\bar{\alpha}^\dagger\alpha$ is the helicity operator.

\subsection*{General ansatz}
The ansatz for the contributions to $\Delta_\nu$ can be taken as
\begin{equation}\label{eq:AnsatzDeltaNuFull}
Y^{r_1\ldots r_k s_1\ldots s_l t_1\ldots t_m u_1\ldots u_n}\alpha_{r_1}^\dagger\ldots\alpha_{r_k}^\dagger\bar{\alpha}_{s_1}^\dagger\ldots\bar{\alpha}_{s_l}^\dagger p_{t_1}\ldots p_{t_m}\bar{p}_{u_1}\ldots \bar{p}_{u_n}
\end{equation}
where summations are understood and the complex conjugate should be added. The $Y$ are symmetric in the $r,s,t,u$ labels separately. However, the transverse momentum structure is dramatically simplified by instead using the combinations $\PP_{ij}$ and $\BPP_{ij}$ (see formulas \eqref{eq:PPandBPPtransversemomenta}). This will be done henceforth.

There is one further point that can be simplified for a generic vertex operator. The oscillator basis always takes the same form apart form the range of the label sums $\{1,2,\ldots,\nu\}$. Introduce the shorthand notation
\begin{align}\label{eq:OscillatorBasisShortA1}
\mathbf{A}_{k\bar{l}}^\dagger&=\alpha_{r_1}^\dagger\ldots\alpha_{r_k}^\dagger\bar{\alpha}_{s_1}^\dagger\ldots\bar{\alpha}_{s_l}^\dagger
\end{align}
with complex (not hermitean) conjugates
\begin{align}\label{eq:OscillatorBasisShortA1cc}
\mathbf{A}_{\bar{k}l}^\dagger&=\bar{\alpha}_{r_1}^\dagger\ldots\bar{\alpha}_{r_k}^\dagger\alpha_{s_1}^\dagger\ldots\alpha_{s_l}^\dagger
\end{align}
There is of course permutational symmetry in the field labels $r_i$ and $s_j$ that is implicit in $\mathbf{A}_{k\bar{l}}^\dagger$ and $\mathbf{A}_{\bar{k}l}^\dagger$.

\begin{remark}[Understanding the advantage of this formulation]\label{rem:UAdvantage}
This setup and formalism may seem redundant but it has the advantage that the higher spin fields themselves may be factored out of the computations. Colour ordering issues for odd spin fields does not interfere with the derivations of the equations for the vertex operators. Such questions can be postponed until particular vertices for particular fields need to be written down. All interaction data is maintained by the $Y$-functions in the operators $\Delta$. Furthermore, the dependence on transverse momenta is expressed in powers of the $\PP_{ij}$ and $\BPP_{ij}$ variables. Thus, interaction data is essentially encoded in these integer powers and rational functions of the $p^+_{i}$ momenta, that is, the $\gamma_i$ variables.
\end{remark}

\section{Review of the cubic vertex operator}\label{sec:ReviewCubic}

For the cubic vertex we may use the general ansatz
\begin{equation}\label{eq:AnsatzDeltaCubicYGen}
\Delta_3=\varrho Y^{(k)(l)mn}\left(\mathbf{A}_{k\bar{l}}^\dagger\PP^m\BPP^n+\mathbf{A}_{\bar{k}l}^\dagger\BPP^m\PP^n\right)
\end{equation}
where $Y^{(k)(l)mn}$ are real rational functions of the $\gamma_r$ to be determined and $\varrho$ is a dimensionful coupling factor. The notation $(k)$ and $(l)$ serves as a reminder that the $Y$-functions have full permutational symmetry in the field labels $r_i$ and $s_j$ respectively. The formula should be interpreted such that $k$ and $l$ are summed $k=1,2,\ldots$ and $l\leq k$ and the explicit form is
\begin{equation}\label{eq:ExplicitForm}
Y^{(k)(l)mn}\mathbf{A}_{k\bar{l}}^\dagger=\sum_{r_1\ldots r_k, s_1\ldots s_l}Y^{r_1\ldots r_ks_1\ldots s_lmn}\alpha_{r_1}^\dagger\ldots\alpha_{r_k}^\dagger\bar{\alpha}_{s_1}^\dagger\ldots\bar{\alpha}_{s_l}^\dagger
\end{equation}
Note that $k-l=n-m$ from $j$-rotational invariance in the transverse space. The ansatz is such that $\Delta_\nu$ is real. For $m=n$ (and therefore $k=l$) the two terms in the ansatz are equal. The ansatz is redundant as it stands since terms with $\PP\BPP$ corresponds to field redefinitions of the free hamiltonian. They can be removed by keeping only terms with either $m=0$ or $n=0$. Thus it is enough to use
\begin{equation}\label{eq:AnsatzDeltaCubicYNoFieldRedefs}
\Delta_3=\varrho_3 Y^{(k)(l)n}\left(\mathbf{A}_{k\bar{l}}^\dagger\BPP^n+\mathbf{A}_{\bar{k}l}^\dagger\PP^n\right)
\end{equation}
with $m=0$ and $n=k-l$.

The ansatz for the Lorentz prefactor $\hat{x}_3$ is such that there is one term for each term in $\Delta_3$ but with one factor less of $\BPP$. Correspondingly, the ansatz for the Lorentz prefactor $\hat{\bar{x}}_3$ is such that there is one term for each term in $\Delta_3$ but with one factor less of $\PP$. Thus
\begin{equation}\label{eq:AnsatzDeltaCubicC}
\begin{split}
\hat{x}_3&=a^r x_r+\varrho_3 c^{(k)(l)n}\mathbf{A}_{k\bar{l}}^\dagger\BPP^{n-1}\\
\hat{\bar{x}}_3&=\bar{a}^r \bar{x}_r+\varrho_3\bar{c}^{(k)(l)n}\mathbf{A}_{\bar{k}l}^\dagger\PP^{n-1}
\end{split}
\end{equation}

\subsubsection*{A note on the coupling factors}
 The coupling factor $\varrho_3$ will be suppressed in the ensuing computations only to be reinstated at the end. The notation $\varrho_\nu$ will be used to distinguish coupling factors corresponding to different vertex orders $\nu$. As we will see, the dimension of the coupling factors will depend on $k$ and $l$ so we can write $\varrho_{\nu(k,l)}$ to keep track of this fact.
\subsection{Computation of a differential commutator}\label{subsec:CompDiffCommQubic}

Let us do the commutator \eqref{eq:DifferentialSideA}, that is $\sum_{r=1}^\nu j_r^-\vert V_{1\ldots\nu}\rangle-\sum_{r=1}^\nu h_r\hat{x}_{(\nu)}\vert V_{1\ldots\nu}\rangle$. There are four terms contributing to this commutator and we list and compute them one by one.

\paragraph{Terms from $xh$ :}
These are (see the first formula of \eqref{eq:JminusLorentzGenerators})
\begin{equation}\label{eq:CalculationJMinusHxh}
\begin{split}
\sum_r x_rh_r\vert V\rangle&=\sum_r\big(h_rx_r+[x_r,h_r]\big)\vert V\rangle\\
&=\sum_r h_r\big([x_r,e^\Delta]+e^\Delta{x_r}\big)\vert\varnothing\rangle+i\sum_r\frac{p_r}{\gamma_r}\vert V\rangle\\
&=\sum_r h_r[x_r,e^\Delta]\vert\varnothing\rangle+\sum_r e^\Delta h_rx_r\vert\varnothing\rangle+i\sum_r\frac{p_r}{\gamma_r}\vert V\rangle
\end{split}
\end{equation}
In the last line, the second and the third terms will cancel contributions from the other parts of the commutator \eqref{eq:DifferentialSideA}. To compute the first term, note that $h_r[x_r,\cdot]$ will act on each and every $\PP$ and $\BPP$ in $\Delta$ with the result
\begin{equation}\label{eq:XHonPP}
\sum_{r=1}^3h_r[x_r,\PP]=0\quad\text{ and }\quad \sum_{r=1}^3h_r[x_r,\BPP]=\frac{i}{3}\left(\PP\sum_{r=1}^3\frac{\bar{p}_r}{\gamma_r}+\BPP\sum_{r=1}^3\frac{p_r}{\gamma_r}\right)
\end{equation}
The first term on the last line of \eqref{eq:CalculationJMinusHxh} therefore becomes
\begin{equation}\label{eq:JMinusHxh}
\begin{split}
\frac{in}{3}Y^{(k)(l)n}\left(\PP\sum_{r=1}^3\frac{\bar{p}_r}{\gamma_r}+\BPP\sum_{r=1}^3\frac{p_r}{\gamma_r}\right)\BPP^{n-1}\mathbf{A}_{k\bar{l}}\vert V\rangle
\end{split}
\end{equation}
%

\paragraph{Terms from $ip\frac{\partial}{\partial\gamma}$ :}

These terms are computed using 
\begin{equation}\label{eq:PdGammaonPP}
\sum_{r=1}^3p_r\frac{\partial}{\partial\gamma_r}\PP=0\quad\text{ and }\quad \sum_{r=1}^3p_r\frac{\partial}{\partial\gamma_r}\BPP=\frac{1}{3}\left(\BPP\sum_{r=1}^3\frac{p_r}{\gamma_r}-\PP\sum_{r=1}^3\frac{\bar{p}_r}{\gamma_r}\right)
\end{equation}
and noting the action on the $|\varnothing\rangle$ vacuum
\begin{equation}\label{eq:DGammaOnVacuum}
\sum_r p_r\frac{\partial}{\partial\gamma_r}\Gamma^{-1}\delta({\textstyle\sum_r}\gamma_r)=-\sum_r\frac{p_r}{\gamma_r}\delta({\textstyle\sum_r}\gamma_r)
\end{equation}
Using this we get
\begin{equation}\label{eq:JMinusHpdgamma}
\begin{split}
&i\sum_r p_r\frac{\partial}{\partial\gamma_r}e^\Delta\vert \varnothing\rangle=i\sum_r \bigg(p_r\frac{\partial\Delta}{\partial\gamma_r}\bigg)e^\Delta\vert \varnothing\rangle+i\sum_r e^\Delta p_r\frac{\partial}{\partial\gamma_r}\vert \varnothing\rangle\\
=&i\sum_r p_r\frac{\partial Y^{(k)(l)n}}{{\partial\gamma_r}}\BPP^{n}\mathbf{A}_{k\bar{l}}\vert V\rangle+i\sum_r p_r\frac{\partial Y^{(k)(l)n}}{{\partial\gamma_r}}\PP^{n}\mathbf{A}_{\bar{k}l}\vert V\rangle\\
+&\frac{in}{3}Y^{(k)(l)n}\left(\BPP\sum_{r=1}^3\frac{p_r}{\gamma_r}-\PP\sum_{r=1}^3\frac{\bar{p}_r}{\gamma_r}\right)\BPP^{n-1}\mathbf{A}_{k\bar{l}}\vert V\rangle-i\sum_r\frac{p_r}{\gamma_r}\vert V\rangle
\end{split}
\end{equation}
where the last term cancels the third term from the last line of  \eqref{eq:CalculationJMinusHxh}.

\paragraph{Terms from $-\frac{i}{\gamma}Mp$ :} The annihilators in $M$ act on the creators in $\Delta$ inserting a term $p/\gamma$ for every $\alpha^\dagger$ and a term $-p/\gamma$ for every $\bar{\alpha}^\dagger$. The result is (here we need explicit indices)
\begin{equation}\label{eq:JMinusHMp}
\begin{split}
&-iY^{r_1\ldots r_k s_1\ldots s_ln}\bigg(\frac{p_{r_1}}{\gamma_{r_1}}+\ldots+\frac{p_{r_k}}{\gamma_{r_k}}-\frac{p_{s_1}}{\gamma_{s_1}}-\ldots-\frac{p_{s_l}}{\gamma_{s_l}}\bigg)\BPP^n\mathbf{A}_{k\bar{l}}\vert V\rangle\\
&+iY^{r_1\ldots r_k s_1\ldots s_ln}\bigg(\frac{p_{r_1}}{\gamma_{r_1}}+\ldots+\frac{p_{r_k}}{\gamma_{r_k}}-\frac{p_{s_1}}{\gamma_{s_1}}-\ldots-\frac{p_{s_l}}{\gamma_{s_l}}\bigg)\PP^n\mathbf{A}_{\bar{k}l}\vert V\rangle
\end{split}
\end{equation}

Then there remains to compute the second term in \eqref{eq:DifferentialSideA}. This essentially entails commuting the prefactor $\hat{x}$ through $e^\Delta$. We do it first for the coordinate piece, then for the oscillator piece.

\paragraph{Terms from $a^r x_r$ :} The computation runs as follows
\begin{equation}\label{eq:CalculationJMinusHhax}
\begin{split}
-\sum_t h_t\bigg(\sum_r a_rx_r\bigg)\vert V\rangle)=&-\sum_t h_t e^\Delta\sum_r a_r[x_r,\Delta]\vert\varnothing\rangle-\sum_t e^\Delta h_t\sum_r a_rx_r\vert\varnothing\rangle
\end{split}
\end{equation}
The second term cancels the second term of \eqref{eq:CalculationJMinusHxh} provided we choose all $a_r=1/\nu$. This is because all $x_r$ are equal on the vacuum, a consequence of momentum conservation, or locality in transverse directions. We are left with the first term. It is also zero since $\sum_{r=1}^3a_r[x_r,\BPP]\sim\sum_{r=1}^3\widetilde{\gamma}_r=0$ when all $a_r$ are equal.
%

\paragraph{Terms from $c^{(k)(l)}$ :} These terms commute with everything in the vertex and so just become multiplications. The contribution is
\begin{equation}\label{eq:JMinusHhcalfa}
-c^{(k)(l)n}\BPP^{n-1}\sum_r h_r\mathbf{A}_{k\bar{l}}\vert V\rangle
\end{equation}
where $\sum_rh_r=-\PP\BPP/\Gamma$.

\subsection{The cubic differential equations}\label{subseq:CubicDiffEq}

Collecting the non-cancelling terms from \eqref{eq:JMinusHxh}, \eqref{eq:JMinusHpdgamma}, \eqref{eq:JMinusHMp} and \eqref{eq:JMinusHhcalfa}, we get two equations, one with oscillator basis $\mathbf{A}_{k\bar{l}}\vert V\rangle$

\begin{equation}\label{eq:HJ-Differential1CubicSimplified}
\begin{split}
\Big[i&\sum_r p_r\frac{\partial Y^{(k)(l)n}}{{\partial\gamma_r}}\BPP^n+\frac{2in}{3}Y^{(k)(l)n}\BPP^n\sum_r\frac{p_r}{\gamma_r}+\Gamma^{-1} c^{(k)(l)n}\PP\BPP^{n}\\
&-iY^{r_1\ldots r_k s_1\ldots s_ln}\left(\frac{p_{r_1}}{\gamma_{r_1}}+\ldots+\frac{p_{r_k}}{\gamma_{r_k}}-\frac{p_{s_1}}{\gamma_{s_1}}-\ldots-\frac{p_{s_l}}{\gamma_{s_l}}\right)\BPP^n\Big]\mathbf{A}_{k\bar{l}}\vert V\rangle=0
\end{split}
\end{equation}
and one with oscillator basis $\mathbf{A}_{\bar{k}l}\vert V\rangle$
\begin{equation}\label{eq:HJ-Differential2CubicSimplified}
\begin{split}
\Big[i&\sum_r p_r\frac{\partial Y^{(k)(l)n}}{{\partial\gamma_r}}\PP^n+iY^{r_1\ldots r_k s_1\ldots s_l}\left(\frac{p_{r_1}}{\gamma_{r_1}}+\ldots+\frac{p_{r_k}}{\gamma_{r_k}}-\frac{p_{s_1}}{\gamma_{s_1}}-\ldots-\frac{p_{s_l}}{\gamma_{s_l}}\right)\PP^n\Big]\mathbf{A}_{\bar{k}l}\vert V\rangle=0
\end{split}
\end{equation}
Expanding out these equations, we get two equations for each concrete index combination $r_1\ldots r_k s_1\ldots s_l$. Since there is no source for the cubic differential, factors of $\mathbb{P}$ and $\mathbb{\bar{P}}$ can now be factored out. This finally gives

\begin{equation}\label{eq:HJ-Differential1CubicFactoredP}
\begin{split}
i&\sum_r p_r\frac{\partial Y^{(k)(l)n}}{{\partial\gamma_r}}+\frac{2in}{3}Y^{(k)(l)n}\sum_r\frac{p_r}{\gamma_r}+\Gamma^{-1} c^{(k)(l)n}\PP\\
&-iY^{r_1\ldots r_k s_1\ldots s_ln}\left(\frac{p_{r_1}}{\gamma_{r_1}}+\ldots+\frac{p_{r_k}}{\gamma_{r_k}}-\frac{p_{s_1}}{\gamma_{s_1}}-\ldots-\frac{p_{s_l}}{\gamma_{s_l}}\right)=0
\end{split}
\end{equation}
and
\begin{equation}\label{eq:HJ-Differential2CubicFactoredP}
\begin{split}
i&\sum_r p_r\frac{\partial Y^{(k)(l)n}}{{\partial\gamma_r}}+iY^{r_1\ldots r_k s_1\ldots s_ln}\left(\frac{p_{r_1}}{\gamma_{r_1}}+\ldots+\frac{p_{r_k}}{\gamma_{r_k}}-\frac{p_{s_1}}{\gamma_{s_1}}-\ldots-\frac{p_{s_l}}{\gamma_{s_l}}\right)=0
\end{split}
\end{equation}
Both these differential equations must be satisfied for any cubic vertex. We note that they are linear in transverse momentum. As they stand, they can be easily solved by adding and subtracting them and using the recombination formula \eqref{eq:recombinationFormulaCubic}. In the next section the equations will be solved in a way amenable to generalisation to higher orders.

\subsection{Solution of the cubic differential equations}\label{subsec:SolutionCubicDifferential}
We apply the recombination formula \eqref{eq:recombinationFormulaCubic} to the derivative term to obtain
\begin{equation}\label{eq:ResummedDerivativeY}
\sum_r p_r\frac{\partial Y^{(k)(l)n}}{{\partial\gamma_r}}=\frac{1}{3}\sum_r \gamma_r\frac{\partial Y^{(k)(l)n}}{{\partial\gamma_r}}\sum_{s=1}^3\frac{p_s}{\gamma_s}-\frac{1}{3}\sum_r S_r\frac{\partial Y^{(k)(l)n}}{{\partial\gamma_r}}\PP
\end{equation}
and to the $M$-rotational term
\begin{equation}\label{eq:NonDiagonalYterm}
\begin{split}
&Y^{r_1\ldots r_k s_1\ldots s_ln}\left(\frac{p_{r_1}}{\gamma_{r_1}}+\ldots+\frac{p_{r_k}}{\gamma_{r_k}}-\frac{p_{s_1}}{\gamma_{s_1}}-\ldots-\frac{p_{s_l}}{\gamma_{s_l}}\right)\\
=\frac{1}{3}&Y^{r_1\ldots r_k s_1\ldots s_ln}\left(k\sum_{s=1}^3\frac{p_s}{\gamma_s}-\bigg[\frac{S_{r_1}}{\gamma_{r_1}}+\ldots+\frac{S_{r_k}}{\gamma_{r_k}}\bigg]\PP\right)\\
-\frac{1}{3}&Y^{r_1\ldots r_k s_1\ldots s_ln}\left(l\sum_{s=1}^3\frac{p_s}{\gamma_s}-\bigg[\frac{S_{s_1}}{\gamma_{s_1}}+\ldots+\frac{S_{s_l}}{\gamma_{s_l}}\bigg]\PP\right)
\end{split}
\end{equation}
Inserting these formulas into \eqref{eq:HJ-Differential1CubicFactoredP} and \eqref{eq:HJ-Differential2CubicFactoredP} we can extract linearly independent parts of the differential equations. First, adding and subtraction the equations along the direction $\sum p_s/\gamma_s$ yield
\begin{subequations}\label{eq:ReducedCubicDifferentialEqs1}
\begin{equation}\label{eq:ReducedCubicDifferentialEqs1A}
\sum_r \gamma_r\frac{\partial Y^{(k)(l)n}}{{\partial\gamma_r}}+nY^{(k)(l)n}=0
\end{equation}
\begin{equation}\label{eq:ReducedCubicDifferentialEqs1B}
-(k-l)+n=0
\end{equation}
\end{subequations}
This is at first a surprising result. The first equation is precisely the same equation that follows from the $j^{+-}$ homogeneity constraint (see \cite{AKHB2012a}). However, this is the light-front reflection of the fact that cubic amplitudes are determined by little group scaling (see for instance  \cite{ElvangHuangBook}). The second equation is the helicity balance equation coming from $j$-invariance. Combining we get
\begin{equation}\label{eq:MinimalYequation}
\sum_r \gamma_r\frac{\partial Y^{(k)(l)n}}{{\partial\gamma_r}}=(l-k)Y^{(k)(l)n}
\end{equation}
We then recover the fundamental solutions generating all higher spin cubic interactions
\begin{equation}\label{eq:MinimalYsolution}
Y^{r_1\ldots r_ks_1\ldots s_ln}=\frac{\gamma_{s_1}\ldots\gamma_{s_l}}{\gamma_{r_1}\ldots\gamma_{r_k}}
\end{equation}
As a check, we see that these functions solve the differential equation \eqref{eq:HJ-Differential2CubicFactoredP}. 

Next we extract the equations along the direction $\PP$ from \eqref{eq:HJ-Differential1CubicFactoredP}
\begin{equation}\label{eq:CubicAlongPP1}
\begin{split}
&-\frac{i}{3}\sum_r S_r\frac{\partial Y^{(k)(l)n}}{{\partial\gamma_r}}+\Gamma^{-1} c^{(k)(l)n}\\
&+\frac{i}{3}Y^{r_1\ldots r_k s_1\ldots s_ln}\Big(\frac{S_{r_1}}{\gamma_{r_1}}+\ldots+\frac{S_{r_k}}{\gamma_{r_k}}-\frac{S_{s_1}}{\gamma_{s_1}}-\ldots-\frac{S_{s_l}}{\gamma_{s_l}}\Big)=0
\end{split}
\end{equation}
and from \eqref{eq:HJ-Differential2CubicFactoredP}
\begin{equation}\label{eq:CubicAlongPP2}
\begin{split}
&-\frac{i}{3}\sum_r S_r\frac{\partial Y^{(k)(l)n}}{{\partial\gamma_r}}-\frac{i}{3}Y^{r_1\ldots r_k s_1\ldots s_ln}\Big(\frac{S_{r_1}}{\gamma_{r_1}}+\ldots+\frac{S_{r_k}}{\gamma_{r_k}}-\frac{S_{s_1}}{\gamma_{s_1}}-\ldots-\frac{S_{s_l}}{\gamma_{s_l}}\Big)=0
\end{split}
\end{equation}
These equations yield two different expressions for the $c$-functions

\begin{equation}\label{eq:Cubic_c1}
c^{(k)(l)n}=\frac{2i}{3}\Gamma \sum_r S_r\frac{\partial Y^{(k)(l)n}}{{\partial\gamma_r}}
\end{equation}
and
\begin{equation}\label{eq:Cubic_c2}
\begin{split}c^{(k)(l)n}=-\frac{2i}{3}\Gamma Y^{r_1\ldots r_k s_1\ldots s_ln}\Big(\frac{S_{r_1}}{\gamma_{r_1}}+\ldots+\frac{S_{r_k}}{\gamma_{r_k}}-\frac{S_{s_1}}{\gamma_{s_1}}-\ldots-\frac{S_{s_l}}{\gamma_{s_l}}\Big)
\end{split}
\end{equation}
Thus we see that the $c$-functions are determined by the $Y$-functions.

As regards the cubic coupling factors, we now see that the mass dimension of $\varrho_3$ is $l-k$. In terms of $\kappa$ we have $\varrho_{3(k,l)}\sim\kappa^{k-l}$.

\section{The homogeneous part of the quartic vertex}\label{sec:HomogenousQuartic}
In choosing an ansatz for the quartic vertex, we may consider terms with the following transverse structure $\PP_{12}^m\BPP_{34}^n$, $\BPP_{12}^m\PP_{34}^n$ and $\PP_{12}^m\PP_{34}^n$, $\BPP_{12}^m\BPP_{34}^n$. We will work through an ansatz with the first two types of terms in detail, and only record the final result (section \ref{eq:MoreVertices}) for an ansatz with the last two types of terms.

For the quartic vertex we then try
\begin{equation}\label{eq:AnsatzDeltaQuarticY}
\begin{split}
\Delta_4=\varrho_4Y^{(k)(l)mn}\left(\mathbf{A}_{k\bar{l}}^\dagger\PP_{12}^m\BPP_{34}^n+\mathbf{A}_{\bar{k}l}^\dagger\BPP_{12}^m\PP_{34}^n\right)
\end{split}
\end{equation}
The terms should be interpreted such that $k$ and $l$ are summed over the oscillator bases and $n-m=k-l$. The ansatz for the Lorentz prefactor $\hat{x}_4$ is such that there is one term for each term in $\Delta_4$ but with one factor less of $\BPP_{12}$ or $\BPP_{34}$. Correspondingly, the ansatz for the Lorentz prefactor $\hat{\bar{x}}_4$ is such that there is one term for each term in $\Delta_4$ but with one factor less of $\PP_{12}$ or $\PP_{34}$. Thus the ansatz becomes
\begin{equation}\label{eq:AnsatzDeltaQuarticC}
\begin{split}
\hat{x}_4=a^rx_r+\varrho_4c_{12}^{(k)(l)mn}\mathbf{A}_{k\bar{l}}^\dagger\PP_{12}^{m}\BPP_{34}^{n-1}+\varrho_4c_{34}^{(k)(l)mn}\mathbf{A}_{\bar{k}l}^\dagger\BPP_{12}^{m-1}\PP_{34}^{n}
\end{split}
\end{equation}
and $\hat{\bar{x}}_4$ is the complex conjugate of $\hat{x}_4$. We run through the computation of one of the differential commutators just as for the cubic. The coupling $\varrho_4$ will not be shown.\footnote{For the time being, we take the powers $m$ and $n$ to be positive, but there is nothing in the algebra that follows that prevents them from taking negative values.}

\subsection{Computation of a quartic differential commutator}\label{subsec:CompDiffCommQuartic}

Again there are four terms contributing to this commutator \eqref{eq:DifferentialSideA}.

\paragraph{Terms from $xh$ :}
As for the cubic we get
\begin{equation}\label{eq:CalculationJMinusHxhQuartic}
\sum_{r=1}^4 x_rh_r\vert V\rangle=\sum_{r=1}^4 h_r[x_r,e^\Delta]\vert\varnothing\rangle+\sum_{r=1}^4 e^\Delta h_rx_r\vert\varnothing\rangle+i\sum_{r=1}^4\frac{p_r}{\gamma_r}\vert V\rangle
\end{equation}
Again, the second and the third terms will cancel contributions from the other parts of the commutator. To compute the first term, note that $h_r[x_r,\cdot]$ will act on the each and every $\BPP_{12}$ and $\BPP_{34}$ in $\Delta$ with the result
\begin{align}\label{eq:XHonPPQuartic}
\sum_{r=1}^4h_r[x_r,\BPP_{12}]&=\frac{i}{2}\left(\BPP_{12}\Big(\frac{p_1}{\gamma_1}+\frac{p_2}{\gamma_2}\Big)+\PP_{12}\Big(\frac{\bar{p}_1}{\gamma_1}+\frac{\bar{p}_2}{\gamma_2}\Big)\right)\\
\sum_{r=1}^4h_r[x_r,\BPP_{34}]&=\frac{i}{2}\left(\BPP_{34}\Big(\frac{p_3}{\gamma_3}+\frac{p_4}{\gamma_4}\Big)+\PP_{34}\Big(\frac{\bar{p}_3}{\gamma_3}+\frac{\bar{p}_4}{\gamma_4}\Big)\right)
\end{align}
The first term of \eqref{eq:CalculationJMinusHxhQuartic} therefore becomes
\begin{equation}\label{eq:JMinusHxhQuartic}
\begin{split}
&\frac{in}{2}Y^{(k)(l)mn}\left(\BPP_{34}\Big(\frac{p_3}{\gamma_3}+\frac{p_4}{\gamma_4}\Big)+\PP_{34}\Big(\frac{\bar{p}_3}{\gamma_3}+\frac{\bar{p}_4}{\gamma_4}\Big)\right)\PP_{12}^{m}\BPP_{34}^{n-1}\mathbf{A}_{k\bar{l}}^\dagger\vert V\rangle\\
+&\frac{im}{2}Y^{(k)(l)mn}\left(\BPP_{12}\Big(\frac{p_1}{\gamma_1}+\frac{p_2}{\gamma_2}\Big)+\PP_{12}\Big(\frac{\bar{p}_1}{\gamma_1}+\frac{\bar{p}_2}{\gamma_2}\Big)\right)\BPP_{12}^{m-1}\PP_{34}^{n}\mathbf{A}_{\bar{k}l}^\dagger\vert V\rangle
\end{split}
\end{equation}
%

\paragraph{Terms from $ip\frac{\partial}{\partial\gamma}$ :}
As for the cubic we get
\begin{equation}\label{eq:CalculationJMinusHpdgammaQuartic}
\begin{split}
i\sum_r p_r\frac{\partial}{\partial\gamma_r}e^\Delta\vert \varnothing\rangle&=i\sum_r \bigg(p_r\frac{\partial\Delta}{\partial\gamma_r}\bigg)e^\Delta\vert \varnothing\rangle-i\sum_r\frac{p_r}{\gamma_r}\vert V\rangle
\end{split}
\end{equation}
The second term cancels the third term from the last line of  \eqref{eq:CalculationJMinusHxhQuartic}. The surviving terms are computed using 
\begin{align}\label{eq:PdGammaonPPQuartic}
\sum_{r=1}^4p_r\frac{\partial}{\partial\gamma_r}\BPP_{12}&=\frac{1}{2}\left(\BPP_{12}\Big(\frac{p_1}{\gamma_1}+\frac{p_2}{\gamma_2}\Big)-\PP_{12}\Big(\frac{\bar{p}_1}{\gamma_1}+\frac{\bar{p}_2}{\gamma_2}\Big)\right)\\
\sum_{r=1}^4p_r\frac{\partial}{\partial\gamma_r}\BPP_{34}&=\frac{1}{2}\left(\BPP_{34}\Big(\frac{p_3}{\gamma_3}+\frac{p_4}{\gamma_4}\Big)-\PP_{34}\Big(\frac{\bar{p}_3}{\gamma_3}+\frac{\bar{p}_4}{\gamma_4}\Big)\right)
\end{align}
Using this, the first term in \eqref{eq:CalculationJMinusHpdgammaQuartic} becomes
\begin{equation}\label{eq:JMinusHpdgammaQuartic}
\begin{split}
&i\sum_{r=1}^4p_r\frac{\partial Y^{(k)(l)mn}}{\partial\gamma_r}\left(\mathbf{A}_{k\bar{l}}^\dagger\PP_{12}^m\BPP_{34}^n+\mathbf{A}_{\bar{k}l}^\dagger\BPP_{12}^m\PP_{34}^n\right)\vert V\rangle\\
+&\frac{in}{2}Y^{(k)(l)mn}\left(\BPP_{34}\Big(\frac{p_3}{\gamma_3}+\frac{p_4}{\gamma_4}\Big)-\PP_{34}\Big(\frac{\bar{p}_3}{\gamma_3}+\frac{\bar{p}_4}{\gamma_4}\Big)\right)\PP_{12}^{m}\BPP_{34}^{n-1}\mathbf{A}_{k\bar{l}}^\dagger\vert V\rangle\\
+&\frac{im}{2}Y^{(k)(l)mn}\left(\BPP_{12}\Big(\frac{p_1}{\gamma_1}+\frac{p_2}{\gamma_2}\Big)-\PP_{12}\Big(\frac{\bar{p}_1}{\gamma_1}+\frac{\bar{p}_2}{\gamma_2}\Big)\right)\BPP_{12}^{m-1}\PP_{34}^{n}\mathbf{A}_{\bar{k}l}^\dagger\vert V\rangle
\end{split}
\end{equation}
It is clear from the expressions \eqref{eq:JMinusHxhQuartic} and \eqref{eq:JMinusHpdgammaQuartic} that half the terms will add and half the terms will cancel.

\paragraph{Terms from $-\frac{i}{\gamma}Mp$ :} The annihilators in $M$ act on the creators in $\Delta$ inserting a term $p/\gamma$ for every $\alpha^\dagger$ and a term $-p/\gamma$ for every $\bar{\alpha}^\dagger$. The result is
\begin{equation}\label{eq:JMinusHMpQuartic}
\begin{split}
&-iY^{r_1\ldots r_ks_1\ldots s_lmn}\left(\frac{p_{r_1}}{\gamma_{r_1}}+\ldots+\frac{p_{r_k}}{\gamma_{r_k}}-\frac{p_{s_1}}{\gamma_{s_1}}-\ldots-\frac{p_{s_l}}{\gamma_{s_l}}\right)\PP_{12}^m\BPP_{34}^n\mathbf{A}_{k\bar{l}}^\dagger\vert V\rangle\\
&+iY^{r_1\ldots r_ks_1\ldots s_lmn}\left(\frac{p_{r_1}}{\gamma_{r_1}}+\ldots+\frac{p_{r_k}}{\gamma_{r_k}}-\frac{p_{s_1}}{\gamma_{s_1}}-\ldots-\frac{p_{s_l}}{\gamma_{s_l}}\right)\BPP_{12}^m\PP_{34}^n\mathbf{A}_{\bar{k}l}^\dagger\vert V\rangle
\end{split}
\end{equation}

Then there remains, as for the cubic, to compute the second term in \eqref{eq:DifferentialSideA}. The prefactor $\hat{x}$ shall be commuted  through $e^\Delta$. We do it first for the coordinate piece, then for the oscillator piece.

\paragraph{Terms from $a^r x_r$ :} Again we have
\begin{equation}\label{eq:CalculationJMinusHhax}
\begin{split}
-\sum_t h_t\bigg(\sum_r a_rx_r\bigg)\vert V\rangle)=&-\sum_t h_t e^\Delta\sum_r a_r[x_r,\Delta]\vert\varnothing\rangle-\sum_t e^\Delta h_t\sum_r a_rx_r\vert\varnothing\rangle
\end{split}
\end{equation}
As before, the second term cancels the second term of \eqref{eq:CalculationJMinusHxhQuartic} provided we choose all $a_r=1/\nu=1/4$. We are left with the first term. It was zero for the cubic, but not here. It is computed using
\begin{equation}\label{eq:aXonBPP}
\sum_{r=1}^4 a_r[x_r,\BPP_{ij}]=\frac{i}{4}(\gamma_i-\gamma_j)
\end{equation}
assuming all $a_r=1/4$. The terms are
\begin{equation}
\begin{split}
-&\frac{in}{4}Y^{(k)(l)mn}(\gamma_3-\gamma_4)\PP_{12}^{m}\BPP_{34}^{n-1}\Big(\sum_{r=1}^4h_r\Big)\mathbf{A}_{k\bar{l}}^\dagger\vert V\rangle\\
-&\frac{im}{4}Y^{(k)(l)mn}(\gamma_1-\gamma_2)\BPP_{12}^{m-1}\PP_{34}^{n}\Big(\sum_{r=1}^4h_r\Big)\mathbf{A}_{\bar{k}l}^\dagger\vert V\rangle
\end{split}
\end{equation}
%

\paragraph{Terms from $c^{(k)(l)}$ :} These terms commute with everything in the vertex and so just become multiplications. The contributions are
\begin{equation}\label{eq:JMinusHhcalfaQuartic}
-\Big(c_{12}^{(k)(l)mn}\mathbf{A}_{k\bar{l}}^\dagger\PP_{12}^{m}\BPP_{34}^{n-1}+c_{34}^{(k)(l)mn}\mathbf{A}_{\bar{k}l}^\dagger\BPP_{12}^{m-1}\PP_{34}^{n}\Big)\sum_{r=1}^4h_r\vert V\rangle
\end{equation}
where the sum over the free hamiltonians is
\begin{equation}\label{eq:QuarticSumFreeHamiltonians}
\sum_{r=1}^4h_r=\frac{\PP_{12}\BPP_{12}}{\gamma_1\gamma_2(\gamma_1+\gamma_2)}+\frac{\PP_{34}\BPP_{34}}{\gamma_3\gamma_4(\gamma_3+\gamma_4)}
\end{equation}
%

\begin{remark}[Understanding the energy sum]\label{rem:UEnergySum}
One would perhaps expect the free hamiltonians to sum to zero, but according to \eqref{eq:QuarticSumFreeHamiltonians} that is not the case. This can be understood as follows. The on-shell condition $p_\mu p^\mu=2(-p^+p^-+p{\bar p})=2(-h\gamma+p{\bar p})=0$ for a single massless particle is solved on the light-front to yield $h=p{\bar p}/\gamma$. But, of course, 3-momentum conservation for $(\gamma, p, {\bar p})$ is not sufficient for energy conservation. That is an independent requirement. Indeed, requiring four-momentum conservation, so that $\mathbf{s}=-(p_1+p_2)^2=-(p_3+p_4)^2$, it is easy to see that $\mathbf{s}=\PP_{12}\BPP_{12}/\gamma_1\gamma_2=\PP_{34}\BPP_{34}/\gamma_3\gamma_4$. But then the expression on the right hand side of \eqref{eq:QuarticSumFreeHamiltonians} is zero (since the $\gamma$'s sum to zero). See also remark \ref{rem:UEquationsII}.
\end{remark}

\subsection{The quartic homogeneous differential equations}\label{subseq:QuarticDiffEq}
Adding the contributions we get two differential equations, one for each type of oscillator basis.

\paragraph{Basis $\mathbf{A}_{k\bar{l}}^\dagger$}
\begin{equation}\label{eq:QuarticEquations3}
\begin{split}
\Bigg(&i\sum_{r=1}^4p_r\frac{\partial Y^{(k)(l)mn}}{\partial\gamma_r}-iY^{r_1\ldots r_ks_1\ldots s_lmn}\left(\frac{p_{r_1}}{\gamma_{r_1}}+\ldots+\frac{p_{r_k}}{\gamma_{r_k}}-\frac{p_{s_1}}{\gamma_{s_1}}-\ldots-\frac{p_{s_l}}{\gamma_{s_l}}\right)\\
&+inY^{(k)(l)mn}\Big(\frac{p_3}{\gamma_3}+\frac{p_4}{\gamma_4}\Big)\Bigg)\PP_{12}^m\BPP_{34}^n\\
&-\Big(\frac{in}{4}Y^{(k)(l)mn}(\gamma_3-\gamma_4)+c_{12}^{(k)(l)mn}\Big)\PP_{12}^{m}\BPP_{34}^{n-1}\sum_{r=1}^4h_r=0
\end{split}
\end{equation}
%
\paragraph{Basis $\mathbf{A}_{\bar{k}l}^\dagger$}
\begin{equation}\label{eq:QuarticEquations4}
\begin{split}
\Bigg(&i\sum_{r=1}^4p_r\frac{\partial Y^{(k)(l)mn}}{\partial\gamma_r}+iY^{r_1\ldots r_ks_1\ldots s_lmn}\left(\frac{p_{r_1}}{\gamma_{r_1}}+\ldots+\frac{p_{r_k}}{\gamma_{r_k}}-\frac{p_{s_1}}{\gamma_{s_1}}-\ldots-\frac{p_{s_l}}{\gamma_{s_l}}\right)\\
&+imY^{(k)(l)mn}\Big(\frac{p_1}{\gamma_1}+\frac{p_2}{\gamma_2}\Big)\Bigg)\BPP_{12}^m\PP_{34}^n\\
&-\Big(\frac{im}{4}Y^{(k)(l)mn}(\gamma_1-\gamma_2)+c_{34}^{(k)(l)mn}\Big)\BPP_{12}^{m-1}\PP_{34}^{n}\sum_{r=1}^4h_r=0
\end{split}
\end{equation}
Or, if we factor out overall powers of transverse momentum bases

\begin{equation}\label{eq:QuarticEquations3}
\begin{split}
\Bigg(&i\sum_{r=1}^4p_r\frac{\partial Y^{(k)(l)mn}}{\partial\gamma_r}-iY^{r_1\ldots r_ks_1\ldots s_lmn}\left(\frac{p_{r_1}}{\gamma_{r_1}}+\ldots+\frac{p_{r_k}}{\gamma_{r_k}}-\frac{p_{s_1}}{\gamma_{s_1}}-\ldots-\frac{p_{s_l}}{\gamma_{s_l}}\right)\\
&+inY^{(k)(l)mn}\Big(\frac{p_3}{\gamma_3}+\frac{p_4}{\gamma_4}\Big)\Bigg)\BPP_{34}-\Big(\frac{in}{4}Y^{(k)(l)mn}(\gamma_3-\gamma_4)+c_{12}^{(k)(l)mn}\Big)\sum_{r=1}^4h_r=0
\end{split}
\end{equation}
\begin{equation}\label{eq:QuarticEquations4}
\begin{split}
\Bigg(&i\sum_{r=1}^4p_r\frac{\partial Y^{(k)(l)mn}}{\partial\gamma_r}+iY^{r_1\ldots r_ks_1\ldots s_lmn}\left(\frac{p_{r_1}}{\gamma_{r_1}}+\ldots+\frac{p_{r_k}}{\gamma_{r_k}}-\frac{p_{s_1}}{\gamma_{s_1}}-\ldots-\frac{p_{s_l}}{\gamma_{s_l}}\right)\\
&+imY^{(k)(l)mn}\Big(\frac{p_1}{\gamma_1}+\frac{p_2}{\gamma_2}\Big)\Bigg)\BPP_{12}-\Big(\frac{im}{4}Y^{(k)(l)mn}(\gamma_1-\gamma_2)+c_{34}^{(k)(l)mn}\Big)\sum_{r=1}^4h_r=0
\end{split}
\end{equation}
%

\subsection{Solution of the quartic homogeneous differential equations}\label{subsec:SolutionQuarticDifferential}
To find solutions to these equations we have to diagonalise them in the basis \eqref{eq:NewQuarticBasis}. We do it in two steps to bring out the essence of the procedure (this will be generalisable to higher order vertices). The sum over the free hamiltonians is rewritten using equation \eqref{eq:QuarticSumFreeHamiltonians}. The recombination formula \eqref{eq:recombinationFormulaQuartic3Schannel} is first applied to the derivative terms. This yields (we also drop the indices on the $Y$ functions where possible without losing information, writing $Y=Y^{(k)(l)mn}$ and drop explicit summation signs)

\begin{equation}\label{eq:QuarticEquations5}
\begin{split}
\Bigg[&\frac{i}{4}\gamma_r\frac{\partial Y}{\partial\gamma_r}\sum_{s=1}^4\frac{p_s}{\gamma_s}-\frac{i}{4}\left(S_{12,r}\frac{\partial Y}{\partial\gamma_r}\right)\PP_{12}-\frac{i}{4}\left(S_{34,r}\frac{\partial Y}{\partial\gamma_r}\right)\PP_{34}\\
&-iY^{r_1\ldots r_ks_1\ldots s_l}\left(\frac{p_{r_1}}{\gamma_{r_1}}+\ldots+\frac{p_{r_k}}{\gamma_{r_k}}-\frac{p_{s_1}}{\gamma_{s_1}}-\ldots-\frac{p_{s_l}}{\gamma_{s_l}}\right)+inY\Big(\frac{p_3}{\gamma_3}+\frac{p_4}{\gamma_4}\Big)\Bigg]\BPP_{34}\\
&-\Big(\frac{in}{4}Y(\gamma_3-\gamma_4)+c_{12}\Big)\left(\frac{\PP_{12}\BPP_{12}}{\gamma_1\gamma_2(\gamma_1+\gamma_2)}+\frac{\PP_{34}\BPP_{34}}{\gamma_3\gamma_4(\gamma_3+\gamma_4)}\right)=0
\end{split}
\end{equation}

\begin{equation}\label{eq:QuarticEquations6}
\begin{split}
\Bigg[&\frac{i}{4}\gamma_r\frac{\partial Y}{\partial\gamma_r}\sum_{s=1}^4\frac{p_s}{\gamma_s}-\frac{i}{4}\left(S_{12,r}\frac{\partial Y}{\partial\gamma_r}\right)\PP_{12}-\frac{i}{4}\left(S_{34,r}\frac{\partial Y}{\partial\gamma_r}\right)\PP_{34}\\
&+iY^{r_1\ldots r_ks_1\ldots s_l}\left(\frac{p_{r_1}}{\gamma_{r_1}}+\ldots+\frac{p_{r_k}}{\gamma_{r_k}}-\frac{p_{s_1}}{\gamma_{s_1}}-\ldots-\frac{p_{s_l}}{\gamma_{s_l}}\right)+imY\Big(\frac{p_1}{\gamma_1}+\frac{p_2}{\gamma_2}\Big)\Bigg]\BPP_{12}\\
&-\Big(\frac{im}{4}Y(\gamma_1-\gamma_2)+c_{34}\Big)\left(\frac{\PP_{12}\BPP_{12}}{\gamma_1\gamma_2(\gamma_1+\gamma_2)}+\frac{\PP_{34}\BPP_{34}}{\gamma_3\gamma_4(\gamma_3+\gamma_4)}\right)=0
\end{split}
\end{equation}
Then we apply the recombination formula to the rest of the non-diagonal terms with the result
\begin{equation}\label{eq:Diagonalise1}
\begin{split}
\frac{p_1}{\gamma_1}+\frac{p_2}{\gamma_2}&=\frac{1}{2}\sum_{s=1}^4\frac{p_s}{\gamma_s}+\frac{1}{2}\frac{\gamma_1-\gamma_2}{\gamma_1\gamma_2(\gamma_1+\gamma_2)}\PP_{12}-\frac{1}{2}\frac{\gamma_3-\gamma_4}{\gamma_3\gamma_4(\gamma_3+\gamma_4)}\PP_{34}\\
\frac{p_3}{\gamma_3}+\frac{p_4}{\gamma_4}&=\frac{1}{2}\sum_{s=1}^4\frac{p_s}{\gamma_s}-\frac{1}{2}\frac{\gamma_1-\gamma_2}{\gamma_1\gamma_2(\gamma_1+\gamma_2)}\PP_{12}+\frac{1}{2}\frac{\gamma_3-\gamma_4}{\gamma_3\gamma_4(\gamma_3+\gamma_4)}\PP_{34}
\end{split}
\end{equation}
and a somewhat unwieldy expression for the terms from the transverse rotations
\begin{equation}\label{eq:Diagonalise2}
\begin{split}
&Y^{r_1\ldots r_ks_1\ldots s_l}\left(\frac{p_{r_1}}{\gamma_{r_1}}+\ldots+\frac{p_{r_k}}{\gamma_{r_k}}-\frac{p_{s_1}}{\gamma_{s_1}}-\ldots-\frac{p_{s_l}}{\gamma_{s_l}}\right)=\\
\frac{1}{4}&Y^{r_1\ldots r_ks_1\ldots s_l}\left(k\sum_{s=1}^4\frac{p_s}{\gamma_s}-\left[\frac{S_{12,r_1}}{\gamma_{r_1}}+\ldots+\frac{S_{12,r_k}}{\gamma_{r_k}}\right]\PP_{12}-\left[\frac{S_{34,r_1}}{\gamma_{r_1}}+\ldots+\frac{S_{34,r_k}}{\gamma_{r_k}}\right]\PP_{34}\right)\\
-\frac{1}{4}&Y^{r_1\ldots r_ks_1\ldots s_l}\left(l\sum_{s=1}^4\frac{p_s}{\gamma_s}-\left[\frac{S_{12,s_1}}{\gamma_{s_1}}+\ldots+\frac{S_{12,s_l}}{\gamma_{s_l}}\right]\PP_{12}-\left[\frac{S_{34,s_1}}{\gamma_{s_1}}+\ldots+\frac{S_{34,s_l}}{\gamma_{s_l}}\right]\PP_{34}\right)
\end{split}
\end{equation}

We can now extract linearly independent parts of the equations. Inserting \eqref{eq:Diagonalise1} and  \eqref{eq:Diagonalise2} into \eqref{eq:QuarticEquations5} and \eqref{eq:QuarticEquations6} and picking out the terms along the directions $\sum\frac{p_s}{\gamma_s}\BPP_{34}$ and $\sum\frac{p_s}{\gamma_s}\BPP_{12}$ respectively, we get
\begin{align}\label{eq:gammaDgamma1a}
\gamma_r\frac{\partial Y}{\partial\gamma_r}-(k-l)Y+2nY=0\\\label{eq:gammaDgamma1b}
\gamma_r\frac{\partial Y}{\partial\gamma_r}+(k-l)Y+2mY=0
\end{align}
from which results
\begin{subequations}\label{eq:gammaDgamma2}
\begin{equation}\label{eq:gammaDgamma2A}
\gamma_r\frac{\partial Y}{\partial\gamma_r}=-(m+n)Y
\end{equation}
\begin{equation}\label{eq:gammaDgamma2B}
k-l=n-m
\end{equation}
\end{subequations}
The first equation is the same homogeneity equation for the $Y$-functions as we had for the cubic vertex. It is the same equation as results from $j^{+-}$ invariance of the vertex. The second equation is the helicity balance of the vertex. \emph{Let us pause and understand this.}

\begin{remark}[Understanding the equations]\label{UEquationsI}
Both for the cubic and the quartic calculation we see that the equations (see \eqref{eq:ReducedCubicDifferentialEqs1A}, \eqref{eq:ReducedCubicDifferentialEqs1B} and \eqref{eq:gammaDgamma2A}, \eqref{eq:gammaDgamma2B}) along the $\sum p_r/\gamma_r$ direction just give back the kinematic restrictions of $\gamma$-homogeneity ($j^{+-}$ invariance) and helicity balance ($j$ invariance). The deformation equations in these directions will remain homogeneous since the source commutator produce no such terms. The dynamical restrictions will lie along the directions of linearly independent powers of $\PP_{ij}$ and $\BPP_{ij}$. For the cubic, no further restrictions occur, as we have seen from the explicit calculations.
\end{remark}

What now remains of the differential equations are the following pieces. From \eqref{eq:QuarticEquations5} we get
\paragraph{Terms $\PP_{12}\BPP_{34}$}
\begin{equation}\label{eq:PP12BPP34_1}
\begin{split}
&-\frac{i}{4}S_{12,r}\frac{\partial Y}{\partial\gamma_r}-\frac{in}{2}\frac{\gamma_1-\gamma_2}{\gamma_1\gamma_2(\gamma_1+\gamma_2)}Y\\
&+\frac{i}{4}\left[\frac{S_{12,r_1}}{\gamma_{r_1}}+\ldots+\frac{S_{12,r_k}}{\gamma_{r_k}}-\frac{S_{12,s_1}}{\gamma_{s_1}}-\ldots-\frac{S_{12,s_l}}{\gamma_{s_l}}\right]Y^{r_1\ldots r_ks_1\ldots s_l}=0
\end{split}
\end{equation}

\paragraph{Terms $\PP_{34}\BPP_{34}$}
\begin{equation}\label{eq:PP34BPP34_1}
\begin{split}
&-\frac{i}{4}S_{34,r}\frac{\partial Y}{\partial\gamma_r}+\frac{in}{2}\frac{\gamma_3-\gamma_4}{\gamma_3\gamma_4(\gamma_3+\gamma_4)}Y-\Big(\frac{in}{4}Y(\gamma_3-\gamma_4)+c_{12}\Big)\frac{1}{\gamma_3\gamma_4(\gamma_3+\gamma_4)}\\
&+\frac{i}{4}\left[\frac{S_{34,r_1}}{\gamma_{r_1}}+\ldots+\frac{S_{34,r_k}}{\gamma_{r_k}}-\frac{S_{34,s_1}}{\gamma_{s_1}}-\ldots-\frac{S_{34,s_l}}{\gamma_{s_l}}\right]Y^{r_1\ldots r_ks_1\ldots s_l}=0
\end{split}
\end{equation}

\paragraph{Terms $\PP_{12}\BPP_{12}$}
\begin{equation}\label{eq:PP12BPP12_1}
\begin{split}
\Big(\frac{in}{4}Y(\gamma_3-\gamma_4)+c_{12}\Big)\frac{1}{\gamma_1\gamma_2(\gamma_1+\gamma_2)}=0
\end{split}
\end{equation}

From \eqref{eq:QuarticEquations6} we get

\paragraph{Terms $\PP_{34}\BPP_{12}$}
\begin{equation}\label{eq:PP34BPP12_2}
\begin{split}
&-\frac{i}{4}S_{34,r}\frac{\partial Y}{\partial\gamma_r}-\frac{im}{2}\frac{\gamma_3-\gamma_4}{\gamma_3\gamma_4(\gamma_3+\gamma_4)}Y\\
&-\frac{i}{4}\left[\frac{S_{34,r_1}}{\gamma_{r_1}}+\ldots+\frac{S_{34,r_k}}{\gamma_{r_k}}-\frac{S_{34,s_1}}{\gamma_{s_1}}-\ldots-\frac{S_{34,s_l}}{\gamma_{s_l}}\right]Y^{r_1\ldots r_ks_1\ldots s_l}=0
\end{split}
\end{equation}

\paragraph{Terms $\PP_{12}\BPP_{12}$}
\begin{equation}\label{eq:PP12BPP12_2}
\begin{split}
&-\frac{i}{4}S_{12,r}\frac{\partial Y}{\partial\gamma_r}+\frac{im}{2}\frac{\gamma_1-\gamma_2}{\gamma_1\gamma_2(\gamma_1+\gamma_2)}Y-\Big(\frac{im}{4}Y(\gamma_1-\gamma_2)+c_{34}\Big)\frac{1}{\gamma_1\gamma_2(\gamma_1+\gamma_2)}\\
&-\frac{i}{4}\left[\frac{S_{12,r_1}}{\gamma_{r_1}}+\ldots+\frac{S_{12,r_k}}{\gamma_{r_k}}-\frac{S_{12,s_1}}{\gamma_{s_1}}-\ldots-\frac{S_{12,s_l}}{\gamma_{s_l}}\right]Y^{r_1\ldots r_ks_1\ldots s_l}=0
\end{split}
\end{equation}

\paragraph{Terms $\PP_{34}\BPP_{34}$}
\begin{equation}\label{eq:PP34BPP34_2}
\begin{split}
\Big(\frac{im}{4}Y(\gamma_1-\gamma_2)+c_{34}\Big)\frac{1}{\gamma_3\gamma_4(\gamma_3+\gamma_4)}=0
\end{split}
\end{equation}

\subsection{Explicit solutions}\label{eq:ExplicitSolution}
Equations \eqref{eq:PP12BPP12_1} and \eqref{eq:PP34BPP34_2} immediately give
\begin{equation}\label{eq:Determine_c_functions}
\begin{split}
c_{12}&=\frac{in}{4}Y(\gamma_3-\gamma_4)\\
c_{34}&=\frac{im}{4}Y(\gamma_1-\gamma_2)
\end{split}
\end{equation}
effectively expressing the dynamical Lorentz generators in terms of the Hamiltonian.

Next adding \eqref{eq:PP12BPP34_1}, \eqref{eq:PP12BPP12_2} and \eqref{eq:PP34BPP34_1}, \eqref{eq:PP34BPP12_2} respectively yield
\begin{equation}\label{S_equations1}
\begin{split}
S_{12,r}\frac{\partial Y}{\partial\gamma_r}=(m-n)\frac{\gamma_1-\gamma_2}{\gamma_1\gamma_2(\gamma_1+\gamma_2)}Y\\
S_{34,r}\frac{\partial Y}{\partial\gamma_r}=(n-m)\frac{\gamma_3-\gamma_4}{\gamma_3\gamma_4(\gamma_3+\gamma_4)}Y
\end{split}
\end{equation}
Expanding the left hand sides and using \eqref{eq:gammaDgamma2A} finally give
\begin{equation}\label{S_equations2}
\begin{split}
\gamma_1\gamma_2\left(\frac{\partial Y}{\partial\gamma_1}-\frac{\partial Y}{\partial\gamma_2}\right)=\frac{m}{2}(\gamma_1-\gamma_2)Y\\
\gamma_3\gamma_4\left(\frac{\partial Y}{\partial\gamma_3}-\frac{\partial Y}{\partial\gamma_4}\right)=\frac{n}{2}(\gamma_3-\gamma_4)Y
\end{split}
\end{equation}

Since the equations are symmetric in $\gamma_1$, $\gamma_2$ and $\gamma_3$, $\gamma_4$ respectively, we can look for solutions depending on sums and products of these variables. Introduce generic variables $x$ and $y$. Then both equations have the form
\begin{equation}\label{S_generic}
xy\left(\frac{\partial}{\partial x}-\frac{\partial}{\partial y}\right)f(x,y)=a(x-y)f(x,y)
\end{equation}
with $a$ an half-integer. First with $z=xy$ and assuming $f(x,y)=f(xy)$ we get
\begin{equation}\label{S_generic_solving}
xy(y-x)f^\prime(z)=a(x-y)f(z)\Rightarrow zf^\prime(z)=-af(z)
\end{equation}
and the solution is $f(z)=Cz^{-a}$ with a coefficient $C$.

Furthermore, with $w=x+y$ and $f(x,y)=f(z,w)$ we still get $zf^\prime(z)=-af(z)$. The $w$ dependence not being fixed by the differential equation. Therefore, a general class of solutions are $f(x,y)=(xy)^{-a}(x+y)^c$ with $c$ free. This is the kind of $p^+$ momentum dependence we would expect from a vertex $Y$-function. Thus we have
\begin{equation}\label{SolutionsS_equationsGeneral}
Y=C(\gamma_1\gamma_2)^{-a}(\gamma_3\gamma_4)^{-b}(\gamma_1+\gamma_2)^{-c}(\gamma_3+\gamma_4)^{-d}
\end{equation}
As noted, the numbers $c$ and $d$ are not fixed by the differential equations but they enter the $\gamma$-homogeneity equation \eqref{eq:gammaDgamma2A} from which results $2(a+b)+c+d=m+n$. But the \eqref{S_equations2} give $2a=m$ and $2b=n$. Combining these constraints yield $c+d=0$. Since $\gamma_3+\gamma_4=-(\gamma_1+\gamma_2)$ this leaves no room for any dependence on $\gamma_1+\gamma_2$ or $\gamma_3+\gamma_4$. 

More generally, equation \eqref{S_generic} can be separated by $f(x,y)=\sum_{m}C_mg_m(x)h_m(y)$. Doing that verifies the solutions $f(x,y)=\rho(x+y)(xy)^{-a}$  where $\rho$ is an arbitrary function of $x+y$. The Y-functions therefore has a multiplicative dependence $\rho_{12}(\gamma_1+\gamma_2)\rho_{34}(\gamma_3+\gamma_4)$ with arbitrary functions $\rho_{12}$ and $\rho_{34}$ of total homogeneity zero. The general solutions to equations \eqref{S_equations2} are therefore given by the functions
\begin{equation}\label{SolutionsS_equations}
Y=\frac{C}{(\gamma_1\gamma_2)^a(\gamma_3\gamma_4)^b}
\end{equation}
of homogeneity degree $-(m+n)=-2(a+b)$. These $Y$-functions solve the homogeneous equations with $m=2a$ and $n=2b$.

\begin{remark}[Understanding the equations]\label{rem:UEquationsII}
At first sight, the non-conservation of the free hamiltonians $\sum_r h_r\neq0$, may seem to be a nuisance. The equations would certainly simplify if we had $\sum_r h_r=0$, or rather, if we did not see the terms along the direction $\sum_r h_r$ at all. Metsaev encounters this problem in \cite{MetsaevQuartic2} (see page 2417) when trying to solve the non-homogeneous equations for the vertices. One could consider (as is done in the paper) $\sum_r h_r=0$ an \emph{energy surface} and look for solutions that are non-singular on this surface, the non-singularity requirement being prompted by the need to invert $\sum_r h_r$.

Now we see that the problem reduces to a question of using a suitable basis in the space of field momenta $\{p_1,p_2,p_3,p_4\}$, namely, $\{\sum_{s=1}^4\frac{p_s}{\gamma_s},\; \PP_{12},\; \PP_{34}\}$. In this basis,  $\sum_r h_r$ simply points along the directions $\PP_{12}\BPP_{12}$ and $\PP_{34}\BPP_{34}$ as in the formula \eqref{eq:QuarticSumFreeHamiltonians}.

On the other hand, going back to equations \eqref{eq:QuarticEquations5} and \eqref{eq:QuarticEquations6} and specialising them to the energy surface, we see that the last term of each equation drop out. This will yield precisely the same equations for the $Y$-functions as off the energy surface, but we get no equations connecting the $c_{12}$- and $c_{34}$-functions to the $Y$-functions.
\end{remark}

\subsection{Consistency checks}\label{eq:CollectingConsistency}
So far the equations resulting from the homogeneous differential equations that we have discussed are \eqref{eq:gammaDgamma2A}, \eqref{eq:gammaDgamma2B} ($\gamma$-homogeneity and helicity balance), \eqref{eq:Determine_c_functions} (equations for the Lorentz prefactors) and \eqref{S_equations2} for the $Y$-functions. There is however more independent information to check. Going back and subtracting \eqref{eq:PP12BPP34_1}, \eqref{eq:PP12BPP12_2} and \eqref{eq:PP34BPP34_1}, \eqref{eq:PP34BPP12_2} respectively yield
\begin{equation}\label{Jrotation_equations1}
\begin{split}
\left[\frac{S_{12,r_1}}{\gamma_{r_1}}+\ldots+\frac{S_{12,r_k}}{\gamma_{r_k}}-\frac{S_{12,s_1}}{\gamma_{s_1}}-\ldots-\frac{S_{12,s_l}}{\gamma_{s_l}}\right]Y^{r_1\ldots r_ks_1\ldots s_l}&=(m+n)\frac{\gamma_1-\gamma_2}{\gamma_1\gamma_2(\gamma_1+\gamma_2)}Y\\
\left[\frac{S_{34,r_1}}{\gamma_{r_1}}+\ldots+\frac{S_{34,r_k}}{\gamma_{r_k}}-\frac{S_{34,s_1}}{\gamma_{s_1}}-\ldots-\frac{S_{34,s_l}}{\gamma_{s_l}}\right]Y^{r_1\ldots r_ks_1\ldots s_l}&=-(m+n)\frac{\gamma_3-\gamma_4}{\gamma_3\gamma_4(\gamma_3+\gamma_4)}Y
\end{split}
\end{equation}
The solutions \eqref{SolutionsS_equations} satisfy these equations, but there is slight ambiguity to sort out. Shall we consider $\gamma_1\gamma_2$ and $\gamma_3\gamma_4$ as $r_i$ and $s_j$ indices respectively or vice versa?\footnote{This ambiguity does not occur for the cubics since in that case the $\gamma$'s are distinguished as to whether they occur in the denominator or the numerator.} Depending on which choice we make we get the right or wrong sign on the right hand sides of \eqref{Jrotation_equations1}. It turns out that $\gamma_1\gamma_2$ corresponds to the $s_j$ indices (i.e. oscillators $\bar{\alpha}^\dagger$) and $\gamma_3\gamma_4$ to the $r_i$ indices (i.e. oscillators $\alpha^\dagger$). Thus, taking $k=n=2b$ and $l=m=2a$, we check explicitly the first equation of \eqref{Jrotation_equations1} for a $Y$-function of the form of \eqref{SolutionsS_equations}
\begin{equation}\label{eq:JrotCalculation1}
\begin{split}
&\left[\frac{S_{12,r_1}}{\gamma_{r_1}}+\ldots+\frac{S_{12,r_k}}{\gamma_{r_k}}-\frac{S_{12,s_1}}{\gamma_{s_1}}-\ldots-\frac{S_{12,s_l}}{\gamma_{s_l}}\right]Y^{r_1\ldots r_ks_1\ldots s_l}\\
=&\left[b\frac{S_{12,3}}{\gamma_3}+b\frac{S_{12,4}}{\gamma_4}-a\frac{S_{12,1}}{\gamma_1}-a\frac{S_{12,2}}{\gamma_2}\right]Y\\
=&(2b+2a)\frac{\gamma_1-\gamma_2}{\gamma_1\gamma_2(\gamma_1+\gamma_2)}Y=(m+n)\frac{\gamma_1-\gamma_2}{\gamma_1\gamma_2(\gamma_1+\gamma_2)}Y
\end{split}
\end{equation}
The second equation of \eqref{Jrotation_equations1} works out similarly.

This is indeed very natural because now the vertices can be written elegantly if we use the light-front -- spinor helicity on-shell dictionary. With
\begin{equation}\label{eq:LFSpinorHelicityDic}
\langle ij\rangle=\langle p_ip_j\rangle=\sqrt{2}\frac{\PP_{ij}}{\sqrt{\gamma_i\gamma_j}}\quad\text{and}\quad[ij]=[ p_ip_j]=-\sqrt{2}\frac{\BPP_{ij}}{\sqrt{\gamma_i\gamma_j}}
\end{equation}
we get
\begin{equation}\label{eq:SpinorHelicityFormVertex1}
\frac{\PP_{12}^m\BPP_{34}^n}{(\gamma_1\gamma_2)^a(\gamma_3\gamma_4)^b}\sim\langle12\rangle^m[34]^n
\end{equation}
%

\subsection{More kinds of vertices}\label{eq:MoreVertices}
It remains to investigate vertex terms with transverse structure $\PP_{12}^m\PP_{34}^n$ and $\BPP_{12}^m\BPP_{34}^n$. Here we will not got through the detailed calculations, but only write down the final differential equations for the $\gamma$ structure and contrast to the terms with $\PP_{12}^m\BPP_{34}^n$ and $\BPP_{12}^m\PP_{34}^n$ treated above. Here we use the ansatz (not showing coupling factors)
\begin{align}\label{eq:AnsatzDeltaQuartic_PP12PP34}
\Delta_4&=Y^{(k)(l)mn}\left(\mathbf{A}_{\bar{k}l}^\dagger\PP_{12}^m\PP_{34}^n+\mathbf{A}_{k\bar{l}}^\dagger\BPP_{12}^m\BPP_{34}^n\right)\\
\hat{x}_4&=a^r x_r+\left(c_{12}^{(k)(l)mn}\BPP_{12}^m\BPP_{34}^{n-1}+c_{34}^{(k)(l)mn}\BPP_{12}^{m-1}\BPP_{34}^{n}\right)\mathbf{A}_{k\bar{l}}^\dagger
\end{align}
where $k-l=m+n$.

We get the following equations
\begin{subequations}\label{eq:PP12pp34equations}
\begin{equation}\label{eq:PP12pp34equationsA}
\begin{cases}
\gamma_r\frac{\partial Y}{\partial\gamma_r}+(m+n)Y=0\\
k=l+m+n=0
\end{cases}
\end{equation}
\begin{equation}\label{eq:PP12pp34equationsB}
\frac{im}{4}Y(\gamma_1-\gamma_2)+c_{34}=0
\end{equation}
\begin{equation}\label{eq:PP12pp34equationsC}
\frac{in}{4}Y(\gamma_3-\gamma_4)+c_{12}=0
\end{equation}
\begin{equation}\label{eq:PP12pp34equationsD}
\left(S_{12,r}\frac{\partial Y}{\partial\gamma_r}\right)-(m-n)\frac{\gamma_1-\gamma_2}{\gamma_1\gamma_2(\gamma_1+\gamma_2)}Y=0
\end{equation}
\begin{equation}\label{eq:PP12pp34equationsE}
\left(S_{34,r}\frac{\partial Y}{\partial\gamma_r}\right)+(m-n)\frac{\gamma_3-\gamma_4}{\gamma_3\gamma_4(\gamma_3+\gamma_4)}Y=0
\end{equation}
\end{subequations}

These equations are the same as for vertices with $\PP_{12}^m\BPP_{34}^n$ and $\BPP_{12}^m\PP_{34}^n$ structure with just one small difference. The relation between $m$, $n$, $k$ and $l$ is different as shown in \eqref{eq:PP12pp34equationsA}. This actually forces $l=0$ since with the $Y$-functions of \eqref{SolutionsS_equations} with all $\gamma$'s in the denominator we must have $k+l=m+n$.

There are also the equations corresponding to the equations of section \ref{eq:CollectingConsistency}.
\begin{equation}\label{Jrotation_equations2}
\begin{split}
\left[\frac{S_{12,r_1}}{\gamma_{r_1}}+\ldots+\frac{S_{12,r_k}}{\gamma_{r_k}}\right]Y^{r_1\ldots r_ks_1\ldots s_l}&=(n-m)\frac{\gamma_1-\gamma_2}{\gamma_1\gamma_2(\gamma_1+\gamma_2)}Y\\
\left[\frac{S_{34,r_1}}{\gamma_{r_1}}+\ldots+\frac{S_{34,r_k}}{\gamma_{r_k}}\right]Y^{r_1\ldots r_ks_1\ldots s_l}&=(m-n)\frac{\gamma_3-\gamma_4}{\gamma_3\gamma_4(\gamma_3+\gamma_4)}Y
\end{split}
\end{equation}
For a solution of the form \eqref{SolutionsS_equations} we get for the first equation
\begin{equation}\label{eq:JrotCalculation2}
\begin{split}
&\left[\frac{S_{12,r_1}}{\gamma_{r_1}}+\ldots+\frac{S_{12,r_k}}{\gamma_{r_k}}\right]Y^{r_1\ldots r_ks_1\ldots s_l}\\
=&\left[a\frac{S_{12,1}}{\gamma_1}+a\frac{S_{12,2}}{\gamma_2}+b\frac{S_{12,3}}{\gamma_3}+b\frac{S_{12,4}}{\gamma_4}\right]Y\\
=&(2b-2a)\frac{\gamma_1-\gamma_2}{\gamma_1\gamma_2(\gamma_1+\gamma_2)}Y=(n-m)\frac{\gamma_1-\gamma_2}{\gamma_1\gamma_2(\gamma_1+\gamma_2)}Y
\end{split}
\end{equation}
%

\section{Quartic vertices}\label{sec:QuarticVertices}
Before writing down explicit formulas for quartic contact vertices, let us first discuss what to expect. It is known (as many authors have noted) that quartic higher spin interactions contain an infinite number of derivatives. Therefore they would contain arbitrarily high powers of transverse momenta on the light-front. In order to understand this, we can contrast with the situation for cubic vertices where we have the Metsaev bounds.\footnote{Derived in \cite{BBL1987} in four dimensions and generalised to general dimensions by Metsaev.}

These bounds restrict the powers of transverse momenta in the vertices. The origin of the bounds are actually very easy to understand. They come about as a consequence of $j$-invariance and the exclusion of vertices containing powers on $\PP\BPP$. There are then basically two types of interaction terms in the Hamiltonian
\begin{equation}\label{eq:TypesBasicCubicInteractions}
\phi_{\lambda_1}\phi_{\lambda_2}\phi_{\lambda_3}Y\BPP^m\quad\text{and}\quad\phi_{\lambda_1}\phi_{\lambda_2}\bar{\phi}_{\lambda_3}Y\BPP^m
\end{equation}
and the complex conjugates thereof. Here, $Y$ stands for the appropriate function of the $p^+$ momenta for the interaction term at hand.

Then $j$-invariance enforces helicity balance, i.e. $m=\lambda_1+\lambda_2+\lambda_3$ and $m=\lambda_1+\lambda_2-\lambda_3$ respectively and we get bounds on the powers of transverse momenta in the vertices. Had we allowed powers of $\PP\BPP$ we would have had interaction terms 
\begin{equation}\label{eq:TypesPPbarCubicInteractions}
\phi_{\lambda_1}\phi_{\lambda_2}\phi_{\lambda_3}Y\BPP^m(\PP\BPP)^n\quad\text{and}\quad\phi_{\lambda_1}\phi_{\lambda_2}\bar{\phi}_{\lambda_3}Y\BPP^m(\PP\BPP)^n
\end{equation}
with arbitrary powers $n$ since the factors $\PP\BPP$ have helicity zero. The full list of such vertices can be found in \cite{AKHB2012a}. The possibility of interaction terms like these has also been noted in a recent paper \cite{CondeJoungMkrtchyan2016}. Normally they would not be considered as proper interactions as $\PP\BPP/\Gamma$ is the sum of the free Hamiltonians.

A type of quartic interaction term in the Hamiltonian has the following structure
\begin{equation}\label{eq:ExampleBasicQuarticInteraction1}
\phi_{\lambda_1}\phi_{\lambda_2}\bar{\phi}_{\lambda_3}\bar{\phi}_{\lambda_4}Y\PP_{12}^m\BPP_{34}^n
\end{equation}
and the complex conjugates thereof. In this case $j$-invariance requires the balance equation ${\lambda_1}+{\lambda_2}-{\lambda_3}-{\lambda_4}=n-m$ to hold. This can be satisfied with arbitrarily high powers $n$ and $m$. 

The presence of interactions of this type can also be understood considering exchange diagrams with two cubic interactions 
\begin{equation}\label{eq:TypesBasicCubicInteractions2}
\phi_{\lambda_1}\bar{\phi}_{\lambda_2}\bar{\phi}_{s}Y_{R}\PP_{12}^{\lambda_2+s-\lambda_1}\quad\text{and}\quad\phi_{s}\bar{\phi}_{\lambda_3}\phi_{\lambda_4}Y_{L}\BPP_{34}^{\lambda_4+s-\lambda_3}
\end{equation}
with $s$ the exchanged helicity with ''propagator'' $(\frac{\PP_{12}\BPP_{34}}{(\gamma_1+\gamma_2)^2})^{-1}$. Qualitatively, this results in

\begin{equation*}\label{eq:ExampleBasicQuarticInteraction2}
\phi_{\lambda_1}\bar{\phi}_{\lambda_2}\bar{\phi}_{\lambda_3}\phi_{\lambda_4}Y_{R}Y_{L}(\gamma_1+\gamma_2)^2\PP_{12}^{\lambda_2-\lambda_1}\BPP_{34}^{\lambda_4-\lambda_3}\big(\PP_{12}\BPP_{34}\big)^{s-1}
\end{equation*}
As arbitrarily high helicities can be exchanged in the channel, this gives interactions with arbitrarily high powers of transverse momenta. 

\subsection*{Typical quartic contact vertex}
We can now put together the homogeneous vertex operator $\Delta_4$ based on the solutions \eqref{SolutionsS_equations} for the $Y$-functions. Preliminary we write
\begin{equation}\label{eq:Homo_Yfunction_prel}
Y^{(k)(l)mn}=\frac{1}{(\gamma_1\gamma_2)^{m/2}(\gamma_3\gamma_4)^{n/2}}=\frac{1}{(\gamma_1\gamma_2)^{l/2}(\gamma_3\gamma_4)^{k/2}}
\end{equation}
The $k$ indices $r_i$ and $l$ indices $s_j$ must be carried by $\gamma_{r_i}$ and $\gamma_{s_j}$. This forces $k=n$ and $l=m$ (see section \ref{eq:CollectingConsistency}) to be even numbers, and the non-zero $Y$-functions must take the form
\begin{equation}\label{eq:Homo_Yfunction_def}
Y^{(1\ldots1\,2\ldots2)(3\ldots3\,4\ldots4)}=\frac{1}{(\gamma_1\gamma_2)^{l/2}(\gamma_3\gamma_4)^{k/2}}
\end{equation}
where there are $l$ indices 1 and 2 and $k$ indices 3 and 4.

Written as contributions to the quartic $\Delta_4$-operator, the solutions to the homogeneous quartic deformation equations with transverse structure of type $\PP_{12}^m\BPP_{34}^n$, take the form
\begin{equation}\label{eq:HomoQuarticInteractionPP12BPP34}
\Delta_4^{\text{hom}}=\sum_{k, l}C_{kl}\varrho_{4,(k,l)}\frac{\PP_{12}^{k}\BPP_{34}^{l}}{(\gamma_1\gamma_2)^{l/2}(\gamma_3\gamma_4)^{k/2}}\mathbf{A}_{k\bar{l}}^\dagger+c.c.
\end{equation}
where we allow for numerical constants $C_{kl}$ not determined at this level. The coupling factors $\varrho_{4,(k,l)}$ have mass dimension $-(k+l)$. 

Particular quartic interactions for fields $\phi_{\lambda_1}$, $\phi_{\lambda_2}$, $\bar{\phi}_{\lambda_3}$ and $\bar{\phi}_{\lambda_4}$ can be extracted from \eqref{eq:OrderNuIntercationFull} and \eqref{eq:NuVertexFull} by computing
\begin{equation}\label{eq:ExampleBasicQuarticInteraction1}
\langle1234\vert\phi_{\lambda_1}\phi_{\lambda_2}\bar{\phi}_{\lambda_3}\bar{\phi}_{\lambda_4}\bar{\alpha}_1^{\lambda_1}\bar{\alpha}_2^{\lambda_2}\alpha_3^{\lambda_3}\alpha_4^{\lambda_4}\exp\Delta_4^{\text{hom}}\vert\varnothing_{1234}\rangle
\end{equation}
Interactions corresponding to the transverse structure of type $\PP_{12}^m\PP_{34}^n$ can be obtained similarly.

\section{Consequences for the constructibility of quartic amplitudes}\label{sec:Consequences}
The existence of solutions to the homogeneous deformation equations shows that there are quartic vertices independent of the cubic vertices, at least to this order in the light-front Poincar{\'e} algebra. It remains to study these terms at the next order (quintic) to see if they survive with non-zero coefficients.

A special approach to quartic amplitudes for massless fields has been investigated by Benincasa and Cachazo  \cite{BenincasaCachazo2008a} and Benincasa and Conde \cite{BenincasaConde2011a,BenincasaConde2011b} in order to map out the consistent interactions among higher and lower spin particles. They use BCFW \cite{BCF2005,BCFW2005} recursion to build quartic tree amplitudes out of cubic amplitudes.

Benincasa and Cachazo \cite{BenincasaCachazo2008a} introduce two concepts, \emph{constructibility} and \emph{the four-particle test}. A theory is constructible if the four-particle tree level amplitudes can be completely computed from the three-particle amplitudes. The four-particle test amounts to computing a certain amplitude using two different BCFW deformations and requiring the results to be equal. Spin 1 (Yang-Mills) and spin 2 (Gravity) pass this test but the test fails for higher spin in that the dependence of the quartic amplitude on the Mandelstam invariants differ for different deformations (see also \cite{AKHB2016b} where these results are reviewed from a light-front perspective). These results have subsequently been refined in \cite{SchusterToro2009a} and \cite{McGadyRodina2014a}.

In order to throw some light on this issue from the point of view of the homogeneous equations for the quartic $Y$-functions, we start by studying the Yang-Mills cubic vertex. The light-front interaction can be expressed as
\begin{equation*}
g^2f^{abe}f^{cde}\int d^4x\frac{1}{\partial^+}\big(\phi^a\partial^+{\bar\phi}^b\big)\frac{1}{\partial^+}\big(\bar\phi^c\partial^+{\phi}^d\big)
\end{equation*}
where $g$ is Yang-Mills coupling constant. In momentum space this corresponds to $Y$-functions of the form
\begin{equation}\label{eq:YM_Yfunction}
\frac{\gamma_2\gamma_4}{(\gamma_1+\gamma_2)(\gamma_3+\gamma_4)}
\end{equation}
This functions has the right homogeneity (zero: $m=n=0$) but it does not, and should not, satisfy the equations \eqref{S_equations2} as it is instead a solution to the non-homogeneous equations connecting the quartic vertex to the cubic.

The spin $2$ quartic interaction is more complex \cite{BengtssonCederwallLindgren1983,Ananth2008}, but as it has to have the correct dimension and conform to the general light-front structure, we can write down the generic $\PP$, $\BPP$ and $\gamma$ structure
\begin{equation*}\label{eq:GR_Yfunction}
\frac{\PP_{12}\BPP_{34}}{(\gamma_1+\gamma_2)(\gamma_3+\gamma_4)}
\end{equation*}
The $\gamma$-homogeneity is of course correct (the $\gamma$ dependence in $\PP$ and $\BPP$ is already discounted), but the equations \eqref{S_equations2} are not satisfied. 

\subsection*{Understanding non-constructibility}
In order to understand this better, let us turn to arbitrary spin $s$. We have generic contact interactions of the form
\begin{equation}\label{eq:HomoQuarticVertex1}
\frac{\PP_{12}^m\BPP_{34}^n}{(\gamma_1\gamma_2)^{l/2}(\gamma_3\gamma_4)^{k/2}}
\end{equation}
The differential equations \eqref{S_equations2} are satisfied with $n=k$ and $k=l$. Consistent with this, helicity balance ($j$ invariance) requires $k-l=n-m$. Then $\gamma$-homogeneity ($j^{+-}$ invariance) is the satisfied since $k+l=m+n$.

Let us now see what kind of contact terms the vertex operators \eqref{eq:HomoQuarticInteractionPP12BPP34} yield. To get a spin 1 four-point coupling we need $k=l=2$ resulting in powers of $(\PP_{12}\BPP_{34})^2$ of transverse momenta. This is clearly impossible for a pure spin 1 theory with a dimensionless coupling constant. Likewise for spin 2 we need $k=l=4$ resulting in 8 powers of transverse momenta which clearly is not compatible with a pure spin 2 theory. The general spin $s$ contact interaction will take the form
\begin{equation}\label{eq:HomoQuarticVertex2}
L^{4s}\frac{(\PP_{12}\BPP_{34})^{2s}}{(\gamma_1\gamma_2\gamma_3\gamma_4)^{s}}
\end{equation}
Where $L$ is a parameter of mass dimension $-1$. Starting from spin 3 there is at least one new dimensionful coupling constant $\alpha_3$ of mass dimension $-2$, in general $\alpha_s$ of mass dimension $1-s$. Therefore $L^{4s}$ can be proportional to $\alpha_{s'}^2$ so that $s'=2s+1$. A spin 3 homogeneous contact term is therefore at the same level as a spin 7 exchange term. This is not as weird as it may seem. If we, in an higher spin theory of this type, do allow a spin $1$ contact term of the form \eqref{eq:HomoQuarticVertex2}, then that contact interaction is at the same level as a quartic spin $1$ exchange interaction with a spin $3$ field in the channel. So if we allow higher spin fields into lower spin theory -- as presumably is unavoidable in full higher spin theory, then contact terms of the type discussed here may play an important role.

Returning to the discussion about non-constructibility, it is phrased in terms of BCFW recursion but the phenomena of non-constructibility does not depend on the  version of complex amplitude deformation technique. The situation may need some further clarification though. The basic questions are: (1) Under what conditions can on-shell quartic tree amplitudes be reconstructed from the cubic amplitudes (constructibility)? (2) Under what conditions can all on-shell higher order tree amplitudes be reconstructed from the cubic amplitudes (full constructibility)? This has been clearly explained in the paper \cite{CohenElvangKiermaier2011a}. The situation is the following.\footnote{Closely following the original explanation in section 6.1 of the cited paper.}

Consider a theory described by a local Lagrangian with interaction vertices of various orders $n$. An on-shell tree amplitude $A_n$ can only depend on interaction vertices with $m\leq n$ fields. However, if $A_n$ can be computed by an on-shell recursion technique, then it has an expression in terms of lower-point ($m<n$) on-shell amplitudes. That is, $A_n$ can be computed without explicit knowledge of any local $n$-point contact interactions.

In Yang-Mills theory there are valid recursion relations for all amplitudes with $n>3$ external lines and all on-shell amplitudes are completely determined by the cubic vertex. The quartic vertex is not needed for on-shell amplitudes. But it must be included in the Lagrangian to make the off-shell Lagrangian gauge invariant. In the light-cone gauge, it is needed for Poincar{\'e} invariance.

In general, reference \cite{CohenElvangKiermaier2011a} defines an $n$-point interaction $Y$ in the local Lagrangian to be a \emph{dependent interaction} if it is completely determined by lower-point interactions, for example through gauge invariance or other symmetries. On the other hand, they refer to $Y$ as an \emph{independent interaction} if the Lagrangian is gauge-invariant and respects all imposed symmetries without the inclusion of $Y$. Dependent $n$-point interactions should not be required as input for on-shell amplitudes, while the information from independent $n$-point interactions must be supplied directly as it cannot be obtained recursively from on-shell amplitudes with less than $n$ external states. This is the situation that we seem to have for Minkowski higher spin theory as set up on the light-front. 

\section{Results}
The main results of the present paper are
\begin{enumerate}
  \item The detailed working out of the differential commutator (in this particular approach) for the quartic vertex. This is generalisable to arbitrary order.
  \item Showing the existence of homogeneous solutions to the differential equations for the quartic vertex and the explicit derivation of light-front quartic contact interactions that are Poincar{\'e} independent of the cubic interactions, thereby confirming previous light-front results \cite{MetsaevQuartic1,MetsaevQuartic2} and covariant results \cite{Taronna2011a}.
  \item Non-constructibility of quartic higher spin amplitudes via BCFW recursion. Although contact terms of the form \eqref{eq:HomoQuarticVertex1} can be constructed for spin 1 and 2 in the context of higher spin theory as formulated here, terms like these are highly unnatural and certainly not needed for the consistency of the pure spin 1 and spin 2 theories. For higher spin theory such quartic interactions may be needed to evade BCFW based no-go arguments. In this context it is interesting to note that in a concrete tree level quantum calculation, the higher spin ''cubic vertices appear to be inconsistent with the BCFW constructibility condition.'' \cite{Ponomarev2016a}. Clearly these questions deserve further study.
\end{enumerate}

\subsection*{Cautionary remarks}
What we have found here is in accordance with and supports previous analysis of non-constructibility of higher spin amplitudes in references \cite{FotopoulosTsulaia2010a,DempsterTsulaia2012a,Taronna2011a,PonomarevTseytlin2016a}. However, let us end with a caveat as to the further consequences beyond BCFW. The homogeneous contact four-point interactions found here do not involve an infinite number of momentum factors for any particular combination of external spin. In this way they more resemble the fundamental cubic interaction terms. So, although invalidating the basic BCFW constructibility assumption (as discussed above) it is by no means clear that their presence saves flat space higher spin theory from other amplitude based inconsistency arguments. 

In a recent set of papers it has been found -- in various settings -- that higher spin four-point amplitudes may only allow trivial scattering. In conformal higher spin theory \cite{Segal2003ConformalHS} it was -- for instance -- found \cite{JoungNakachTseytlin2016a} that the four-scalar tree-level scattering amplitude with a tower of higher spins in the channel vanishes. See also \cite{BeccariaNakachTseytlin2016a} for further results. Similar results -- amplitudes being delta-distributions rather than analytic functions -- are also reported in the AdS/CFT context in \cite{BekaertErdmengerPonomarevSleight2016a} and \cite{Taronna2016a}.

Nonetheless -- in the present authors opinion -- it is important to press forward and try to settle the question of classical consistency of flat space higher spin theory. This question is also tied to the question of whether there exists an infinite dimensional extension of the Poincar{\'e} (or Lorentz) algebra underpinning such a higher spin theory. For a recent analysis of this, see \cite{SleightTaronna2016a}. See also comments in \cite{AKHB2008a}. Somewhat ironically it may be the very existence of such a huge higher spin symmetry that constrains the scattering amplitudes to be trivial, as discussed in \cite{BeccariaNakachTseytlin2016a,BekaertErdmengerPonomarevSleight2016a}.

\section*{Acknowledgement}\label{sec:AcknowledgementX}
I would like to thank Evgeny Skvortsov for discussions on light-front higher spin at the Aspects of Higher Spin workshop in Munich (May 23 to May 25, 2016) and Dmitry Ponomarev for explaining the logic in the Metsaev papers on the quartic interactions during the Amplitudes 2016 conference in Stockholm (July 4 to July 8, 2016).

\pagebreak

\begin{thebibliography}{10}

\bibitem{PonomarevTseytlin2016a}
D.~Ponomarev and A.~Tseytlin.
\newblock On quantum corrections in higher-spin theory in flat space.
\newblock 2016.
\newblock arXiv:1603.06273.

\bibitem{CondeJoungMkrtchyan2016}
E.~Conde, E.~Joung, and K.~Mkrtchyan.
\newblock Spinor-helicity three-point amplitudes from local cubic interactions.
\newblock 2016.
\newblock arXiv:1605.07402.

\bibitem{Ponomarev2016a}
D.~Ponomarev.
\newblock Off-shell spinor-helicity amplitudes from light-cone deformation
  procedure.
\newblock 2016.
\newblock arXiv:1611.00361.

\bibitem{MetsaevQuartic1}
R.~R. Metsaev.
\newblock Poincar{\'e} invariant dynamics of massless higher spins: Fourth
  order analysis on mass shell.
\newblock {\em Mod. Phys. Lett.}, A6:359, 1991.

\bibitem{MetsaevQuartic2}
R.~R. Metsaev.
\newblock S-matrix approach to massless higher spins theory: Ii. the case of
  internal symmetry.
\newblock {\em Mod. Phys. Lett.}, A6:2411, 1991.

\bibitem{BBL1987}
A.~K.~H. Bengtsson, I.~Bengtsson, and N.~Linden.
\newblock Interacting higher-spin gauge fields on the light front.
\newblock {\em Class. Quant. Grav.}, 4:1333, 1987.

\bibitem{AKHB2012a}
A.~K.~H. Bengtsson.
\newblock Systematics of higher-spin light-front interactions.
\newblock 2012.
\newblock arXiv:1205.6117.

\bibitem{AKHB2016a}
A.~K.~H. Bengtsson.
\newblock Notes on cubic and quartic light-front kinematics.
\newblock 2016.
\newblock arXiv:1604.01974.

\bibitem{Taronna2011a}
M.~Taronna.
\newblock Higher-spin interactions: Four-point functions and beyond.
\newblock {\em J. High Energy Phys.}, 0412:029, 2011.
\newblock arXiv:1107.5843.

\bibitem{BenincasaCachazo2008a}
P.~Benincasa and F.~Cachazo.
\newblock Consistency conditions on the s-matrix of massless particles.
\newblock 2008.
\newblock arXiv:0705.4305.

\bibitem{BenincasaConde2011a}
P.~Benincasa and E.~Conde.
\newblock On the tree-level structure of scattering amplitudes of massless
  particles.
\newblock {\em JHEP}, 11(2011)074, 2011.
\newblock arXiv:1106.0166.

\bibitem{AKHB2016b}
A.~K.~H. Bengtsson.
\newblock Quartic amplitudes for minkowski higher spin.
\newblock 2016.
\newblock Contribution to the Proceedings of the International Workshop on
  Higher Spin Gauge Theories, 4-6 November 2015, Singapore. arXiv:1605.02608.

\bibitem{FotopoulosTsulaia2010a}
A.~Fotopoulos and M.~Tsulaia.
\newblock On the tensionless limit of string theory, off - shell higher spin
  interaction vertices and {BCFW} recursion relations.
\newblock {\em JHEP}, 1011:086, 2010.
\newblock arXiv:1009.0727v3.

\bibitem{DempsterTsulaia2012a}
P.~Dempster and M.~Tsulaia.
\newblock {On the Structure of Quartic Vertices for Massless Higher Spin Fields
  on Minkowski Background}.
\newblock {\em Nucl. Phys. B}, 865:353--375, 2012.
\newblock arXiv:1203.5597.

\bibitem{AKHB2005a}
A.~K.~H. Bengtsson.
\newblock An abstract interface to higher spin gauge field theory.
\newblock {\em J. Math. Phys.}, 46:042312, 2005.
\newblock arXiv:hep-th/0403267.

\bibitem{Vasiliev1999Review}
M.~A. Vasiliev.
\newblock Higher spin gauge theories: Star-product and {A}d{S} space.
\newblock 1999.
\newblock arXiv:hep-th/9910096.

\bibitem{Vasiliev2004Reviewa}
M.~A. Vasiliev.
\newblock Higher spin gauge theories in various dimensions.
\newblock {\em Fortsch. Phys.}, 52:702, 2004.
\newblock arXiv:hep-th/0401177.

\bibitem{Vasiliev2004Reviewb}
M.~A. Vasiliev.
\newblock Higher spin gauge theories in any dimension.
\newblock {\em Comptes Rendus Phys.}, 5:1101, 2004.
\newblock arXiv:hep-th/0409260.

\bibitem{Vasiliev2005a}
M.A. Vasiliev.
\newblock Actions, charges and off-shell fields in the unfolded dynamics
  approach.
\newblock {\em Int.J. Geom. Meth. Mod. Phys.}, 3:37--80, 2006.
\newblock arXiv:hep-th/0504090.

\bibitem{BekaertCnockaertIazeollaVasiliev2005}
C.~Iazeolla X.~Bekaert, S.~Cnockaert and M.~A. Vasiliev.
\newblock Nonlinear higher spin theories in various dimensions.
\newblock In G.~Bonelli R.~Argurio, G.~Barnich and M.~Grigoriev, editors, {\em
  First Solvay Workshop on Higher-Spin Gauge Theories}, pages 132--197.
  Universit�e Libre de Bruxelles, International Solvay Institutes for Physics
  and Chemistry, 2004.
\newblock arXiv:hep-th/0503128v2.

\bibitem{NewStructuresPhysics}
B.~Coecke (ed.).
\newblock {\em New Structures for Physics}.
\newblock Springer, Heidelberg, 2011.

\bibitem{Ananth2012un}
S.~Ananth.
\newblock Spinor helicity structures in higher spin theories.
\newblock {\em J. High Energy Phys.}, 1211:089, 2012.
\newblock arXiv:1209.4960.

\bibitem{HaehnelMcLoughlin2016a}
P.~Haehnel and T.~McLoughlin.
\newblock Conformal higher spin theory and twistor space actions.

\bibitem{ElvangHuangBook}
H.~Elvang and Y-T. Huang.
\newblock {\em Scattering Amplitudes in Gauge Theory and Gravity}.
\newblock Cambridge University Press, 2015.

\bibitem{BenincasaConde2011b}
P.~Benincasa and E.~Conde.
\newblock Exploring the s matrix of massless particles.
\newblock {\em Phys. Rev. D}, 86, 025007, 2011.
\newblock arXiv:1108.3078.

\bibitem{BCF2005}
R.~Britto, F.~Cachazo, and B.~Feng.
\newblock New recursion relations for tree amplitudes of gluons.
\newblock {\em Nucl. Phys. B}, 715:499, 2005.
\newblock arXiv:hep/th-0412308.

\bibitem{BCFW2005}
R.~Britto, F.~Cachazo, B.~Feng, and E.~Witten.
\newblock Direct proof of tree-level recursion relation in yang-mills theory.
\newblock {\em Phys. Rev. Lett.}, 94:181602, 2005.
\newblock arXiv:hep/th-0501052.

\bibitem{SchusterToro2009a}
P.~Schuster and N.~Toro.
\newblock Constructing the tree-level {Y}ang-{M}ills {S}-matrix using complex
  factorization.
\newblock {\em JHEP}, 0906:079, 2009.
\newblock arXiv:0811.3207.

\bibitem{McGadyRodina2014a}
Laurentiu~Rodina David A.~McGady.
\newblock Higher-spin massless {S}-matrices in four-dimensions.
\newblock {\em Phys. Rev. D}, 90:084048, 2009.
\newblock arXiv:1311.2938.

\bibitem{BengtssonCederwallLindgren1983}
I.~Bengtsson, M.~Cederwall, and O.~Lindgren.
\newblock Light-cone actions for gravity and higher spins.
\newblock 1983.
\newblock G{\"o}teborg preprint 83-55.

\bibitem{Ananth2008}
S.~Ananth.
\newblock The quintic interaction vertex in light-cone gravity.
\newblock {\em Phys. Lett.}, B664:219--223, 2008.
\newblock arXiv:0803.1494.

\bibitem{CohenElvangKiermaier2011a}
T.~Cohen, H.~Elvang, and M.~Kiermaier.
\newblock On-shell constructibility of tree amplitudes in general field
  theories.
\newblock {\em JHEP}, 1104:053, 2011.
\newblock arXiv:1010.0257.

\bibitem{Segal2003ConformalHS}
Y.~A. Segal.
\newblock Conformal higher spin theory.
\newblock {\em Nucl. Phys. B}, 664.

\bibitem{JoungNakachTseytlin2016a}
E.~Joung, S.~Nakach, and A.~Tseytlin.
\newblock Scalar scattering via conformal higher spin exchange.
\newblock {\em JHEP}, 1602:125, 2016.
\newblock arXiv:1512.08896.

\bibitem{BeccariaNakachTseytlin2016a}
M.~Beccaria, S.~Nakach, and A.~Tseytlin.
\newblock On triviality of {S}-matrix in conformal higher spin theory.
\newblock {\em JHEP}, 1609:034, 2016.
\newblock arXiv:1607.06379.

\bibitem{BekaertErdmengerPonomarevSleight2016a}
X.~Bekaert, J.~Erdmenger, D.~Ponomarev, and C.~Sleight.
\newblock Bulk quartic vertices from boundary four-point correlators.
\newblock 2016.
\newblock arXiv:1602.08571.

\bibitem{Taronna2016a}
M.~Taronna.
\newblock Pseudo-local theories: A functional class proposal.
\newblock 2016.
\newblock arXiv:1602.0866.

\bibitem{SleightTaronna2016a}
C.~Sleight and M.~Taronna.
\newblock Higher-spin algebras, holography and flat space.
\newblock 2016.
\newblock arXiv:1609.00991.

\bibitem{AKHB2008a}
A.~K.~H. Bengtsson.
\newblock Towards unifying structures in higher spin gauge symmetry.
\newblock {\em SIGMA}, 4:013, 2007.
\newblock arXiv:0802.0479.

\end{thebibliography}

\end{document}